\documentclass[published]{JHEP3} 

\JHEP{00(2012)000}

\JHEPspecialurl{http://jhep.sissa.it/JOURNAL/JHEP3.tar.gz}

\usepackage{epsfig,multicol,bbm}

\newcommand\fverb{\setbox\fverbbox=\hbox\bgroup\verb}
\newcommand\fverbdo{\egroup\medskip\noindent%
			\fbox{\unhbox\fverbbox}\ }
\newcommand\fverbit{\egroup\item[\fbox{\unhbox\fverbbox}]}
\newbox\fverbbox

\title{Fluctuations in strongly coupled cosmologies  }

\author{Silvio A. Bonometto\\ Department of Physics,
  Astronomy Unit, Trieste University, Via Tiepolo 11, I~34143 Trieste,
  Italy -- I.N.F.N., Sezione di Trieste, Via Valerio, 2
  I~34127 Trieste, Italy -- I.N.A.F., Astronomical
  Observatory of Trieste, Via Tiepolo 11, I~34143 Trieste, Italy
   }

\author{Roberto Mainini\\ Department of Physics G.~Occhialini,
Milano--Bicocca University, Piazza della Scienza 3, I~20126 Milano,
Italy}

\received{.....}
\accepted{.....}

\abstract{In the early Universe, a dual component made of coupled CDM
  and a scalar field $\Phi$, if their coupling $\beta > \sqrt{3}/2$,
  owns an attractor solution, making them a stationary fraction of
  cosmic energy during the radiation dominated era. Along the
  attractor, both such components expand $\propto a^{-4}$ and have
  early density parameters $\Omega_{d} = 1/ (4\beta^2)$ and $\Omega_c=
  2\, \Omega_d$ (field and CDM, respectively). In a previous paper it
  was shown that, if a further component, expanding $\propto a^{-3}$,
  breaks such stationary expansion at $z \sim 3$--$5 \times 10^3$,
  cosmic components gradually acquire densities consistent with
  observations. This paper, first of all, considers the case that this
  component is warm. However, its main topic is the analysis of
  fluctuation evolution: out of horizon modes are then determined;
  their entry into horizon is numerically evaluated as well as the
  dependence of Meszaros effect on the coupling $\beta$; finally, we
  compute: (i) transfer function and linear spectral function; (ii)
  CMB $C_l$ spectra. Both are close to standard $\Lambda$CDM models;
  in particular, the former one can be so down to a scale smaller than
  Milky Way, in spite of its main DM component being made of particles
  of mass $<1\, $keV. The previously coupled CDM component, whose
  present density parameter is $\cal O$$(10^{-3})$, exhibits wider
  fluctuations $\delta \rho/\rho$, but approximately
  $\beta$--independent $\delta \rho$ values. We discuss how lower
  scale features of these cosmologies might ease quite a few problems
  that $\Lambda$CDM does not easily solve.  }

\keywords{cosmology: theory, dark matter, dark energy, gravitation;
  methods: numerical.}

\begin{document}

\rm
\section{Introduction}
The cosmic dark components (Dark Matter: DM, and Dark Energy: DE) will
be probably considered the main physical discoveries in the last
decades. Understanding their nature and properties is one of the main
tasks of contemporary astrophysical research. A number of options have
been considered and here we shall assume DE to be a scalar field
$\Phi$, interacting with DM, whose lagrangian includes a {\it
  potential} $V(\Phi)$, which could also be a simple mass term. The
hope to detect its shape through the determination of the DE state
equation $w(a)$ is remote ($a:$ scale factor). The very experiment
{\sc Euclid}\footnote{http://www.euclid-ec.org} \cite{Laureijs:2011mu}
is expected to estimate the $w(a)$ derivative at $z=0$ with an error
$\sim 20 \, \%$ \cite{JB}. Accordingly, although assuming DE to be a
scalar field, no explicit $V(\Phi)$ expression will be taken here,
rather focusing on parameters better constrained by observations.

The rational of allowing for DM--DE interactions is to insure DE a
fresh energy inflow, so keeping it a significant fraction of DM
density, at any redshift \cite{amendola}. A fairly large DM--DE
interaction scale
\begin{equation}
\label{C}
C = {b / m_p} = \sqrt{16 \pi / 3}\, \,  {\beta / m_p} ~,
\end{equation}
allowing DE density to steadily keep $\sim 1\, \%$ of CDM (Cold Dark
Matter), is consistent with data: namely, if neutrinos have a mass,
$\beta$ values up to $\sim 0.15$--0.2 \cite{mmc,pettorino,Xia} agree with
observations.

In this work we shall however debate cosmologies where $C$ is still
greater, allowing DE density to keep a steady fraction of {\it
  radiation} through radiative eras. In a previous paper (\cite{BLVS},
paper I here below) this class of cosmologies was shown to be
consistent with data, at background level.

Our main task, here, is to study fluctuations in these models. Out of
horizon fluctuation modes are determined. The entry of fluctuations in
the horizon is then numerically discussed. In particular we compare
Meszaros effect in these cosmologies with standard $\Lambda$CDM
cosmologies, and, finally, work out the transfer function and the
$C_l$ spectra for these cosmologies.

A fit of these cosmologies with observational data is beyond of the
scopes of this work. We however ascertain that discrepancies with
$\Lambda$CDM, far from creating conflicts with data, could be rather
exploited to ease some problems still open in $\Lambda$CDM
cosmologies.

In paper I we showed that, in strongly coupled (S.C.) cosmologies, the
early expansion is characterized by the presence of a dual component,
made of CDM and a scalar field $\Phi$ exchanging energy, aside of
ordinary radiation ($\gamma$'s \& $\nu$'s: photons \& neutrinos). CDM
and $\Phi$ have constant early density parameters
\begin{equation}
\Omega_c = {1 \over 2 \beta^2} ~,~~~ \Omega_d = 
 {1 \over 4 \beta^2} ~,
\label{omgas}
\end{equation}
provided that $\beta$ is constant.

Quite in general, if we allow for an energy leakage from CDM to the
$\Phi$ field, CDM dilution occurs faster than $\propto a^{-3}$. In
turn, being almost kinetic, the field would dilute $\propto a^{-6}$,
unless continuously fed fresh energy, so that its dilution becomes
less frenetic.  The point is that, if the energy leakage becomes so
strong to yield a CDM dilution $\propto a^{-4}$, this is exactly what
is needed to allow the field to dilute at the same rate, provided that
CDM and $\Phi$ density parameters are those in eq.~(\ref{omgas}). The
radiation density parameter is then $\Omega_r = 1-3/4\beta^2$.

Furthermore, as shown in Paper I, this is a tracker regime: starting
from any initial condition, densities and dilution rates rapidly
settle on the regime (\ref{omgas}), provided that
\begin{equation}
\beta > \sqrt{3}/2~,
\end{equation}
and this cosmic tripartition lasts forever, unless a non--relativistic
uncoupled component acquires a significant density. Let us outline
that it could also be such since ever, e.g. since the end of the
inflationary expansion.

As a matter of fact, in the real world baryons exist, at least, whose
density dilutes $\propto a^{-3}$, and will eventually reach the
radiation level. However, if only baryons are {\it added}, as
non--relativistic component, the above picture fails to meet
observational features. They are however met if a further DM component
exist. In principle, it could be another uncoupled CDM component or
warm DM (WDM), becoming non--relativistic (slightly) before a standard
equality redshift.

The rise of such non--relativistic component also causes a rise of CDM
and $\Phi$ densities in respect to radiation. However, until $\Phi$
keeps kinetic, its density still declines more rapidly than
$a^{-3}$. If the progressive rise of the $\Phi$ field is however such
to cause its transition from kinetic to potential, just a little after
WDM derelativized, the background component densities easily meet the
observational proportions.

Altogether, the basic parameter to be adjusted to obtain this result,
is the epoch when, in the energy density of the field
\begin{equation}
\rho_d = {\dot \Phi^2 \over 2 a^2} + V(\Phi)
\label{rhod}
\end{equation}
the latter term will begin to prevail on the former one. This
expression assumes a metric
\begin{equation}
ds^2 = a^2(\tau)(d\tau^2 - d\ell^2)
\label{metric}
\end{equation}
with $\tau$ being the conformal time. Let us also remind that
the $\Phi$ pressure then reads
\begin{equation}
p_d = {\dot \Phi^2 \over 2 a^2} - V(\Phi)
\label{rhod}
\end{equation}
and its state parameter $w(a) = p_d/\rho_d$.

Let us notice that any coupled--DE theory (apart some peculiar
low--$\beta$ cases, when the self--interaction potential is so strong
to make the coupling almost negligible) requires $\Phi$ to be
initially kinetic, then turning to potential at a suitably tuned
redshift. In the literature, this transition has been studied by using
different expressions of the potential $V(\Phi)$, e.g. Ratra--Peebles
\cite{RP} or SUGRA \cite{SUGRA} expressions. Although the detailed
evolution of the DE state parameter $w(a)$, during the transition,
exhibits some potential dependence, the epoch when the transition
takes place is largely independent from the potential shape.

Let us also notice that, for coupled--DE theories, any preference
granted to tracker potentials is unjustified.  As a matter of fact,
initial conditions however assume that the field is purely kinetic,
its value being therefore independent from $V(\Phi)$. The parameters
in the potential are then tuned to enable a kinetic--potential
transition at a suitable epoch. This could equally be done with any
potential expression, e.g. by taking $V(\Phi) = m^2 \Phi^2$ or a
polinomial including higher powers of $\Phi$.

Accordingly, we keep here to the approach of paper I, just assuming a
parametric expression fixing the shape for the $w(a)$ transition from
+1 (at early times) to -1 (close to $z=0$). The expression depends
from a single parameter ($\epsilon$), and the effects of varying
$\epsilon$ mimic changes in the $V(\Phi)$ espression. The expression
chosen here to follow the transition is slightly different from Paper
I, for a specific reason debated in the sequel.

Before concluding this Section let us finally remind that the cosmic
components, in S.C. cosmologies, are the standard ones, apart of DM
which is twofold: early DM is coupled, while an uncoupled component
could be, e.g., WDM.

Also the plan of the paper is essentially twofold. A former part is
devoted to deepening the background picture, taking WDM as uncoupled
DM component. As a matter of fact, this class of cosmologies appears
far more appealing if such upgrade is made. But, as previously
outlined, detailed fits with data are delayed to further work. The
latter part of the paper is then devoted to the study of density
fluctuation evolution.  This will be done by using two numerical
programs: (i) A simple program, solving a set of 11 coupled
differential equations, will enable us to follow closely the physical
features of these models. (ii) We shall then present results obtained
by using a suitably modified version of CMBFAST.

\rm

\section{CDM--DE coupling}
Coupled CDM and DE were considered by several authors
\cite{amendola,mmc,ellis,spain,others}. The point is that the stress--energy
tensors of CDM and DE must fulfill the pseudo--conservation equation
\begin{equation}
\label{eq1}
{T_{(c)}}^\nu_{\mu; \nu} + {T_{(d)}}^\nu_{\mu; \nu} = 0~,
\end{equation}
but there is no direct evidence that (for CDM) $ {T_{(c)}}^\nu_{\mu;
  \nu} = 0$ and/or (for DE) ${T_{(d)}}^\nu_{\mu; \nu} = 0~, $
separately. Following Paper I, here we assume that
\begin{equation}
\label{eq2}
  {T_{(c)}}^\nu_{\mu;\nu} = - CT_{(c)} \Phi_{;\mu} ~,~~~~~
  {T_{(d)}}^\nu_{\mu; \nu} = CT_{(c)} \Phi_{;\mu}
\end{equation}
which is not the only possible option (see, e.g., \cite{spain}), but
is the one considered first, aiming to ease the paradox of DE being
significant only in the present epoch. In a FRW frame, the equations
(\ref{eq2}) yield
\begin{equation}
\label{c1}
\dot \Phi_1 + \tilde w{\dot a \over a} \Phi_1
= {1+w \over 2} C \rho_c a^2~,~~~~~
\dot \rho_c + 3{\dot a \over a} \rho_c = -C \Phi_1 \rho_c~.
\end{equation}
Here 
\begin{equation}
\Phi_1 = \dot \Phi~, ~~~~~~~
2\tilde w = 1+3w-d\log (1+w)/d\log a~,
\end{equation}
and this formulation is equivalent, e.g., to \cite{amendola}, if
$w(a)$ (state equation of DE) is assigned, instead of the
self--interaction potential of the field $V(\Phi)$. 
To obtain  eq.~(\ref{c1}), as in Paper I, we used the expression
\begin{equation}
\label{V'}
(1+w)\, a^2 V'(\Phi) = \dot \Phi_1 (1-w) - 2 {\dot a \over a} \Phi_1
(1+w-\tilde w)
\end{equation}
for the potential derivative.

The latter eq.~(\ref{c1}) has then the formal integral
\begin{equation}
\rho_c = \rho_{i,c} \left(a_i \over a \right)^3
\exp\left(-C\int_{\tau_i}^\tau d\tau \Phi_1 \right)~,
\label{formal}
\end{equation}
$\tau_i$ being a reference time when CDM density is $\rho_{i,c}$ and
the scale factor is $a_i= a(\tau_i)$.  If this expression for $\rho_c$
is then replaced in the former eq.~(\ref{c1}), we obtain a
trascendental differential equation, quite hard to integrate. 
In Paper I, we however outlined that, if we make the ansatz
\begin{equation}
\label{phi1}
\Phi_1 = \alpha~ {m_p / \tau}~,
\end{equation}
it is $ -C\int_{\tau_i}^\tau d\tau~ \Phi_1 = \ln\left(\tau_i /
\tau \right)^{\alpha b}~, $ so that $ \rho_c = \rho_{i,c} \left(a_i
/ a \right)^{3+\alpha b}$ and, by replacing the expression
(\ref{phi1}) in the former eq.~(\ref{c1}), we obtain
\begin{equation}
(\tilde w -1) \alpha {m_p \over a^2 \tau^2} = {1+w \over 2} {b 
\over m_p} \rho_{i,c} \left(a_i \over a \right)^{3+\alpha b}~,
\label{same}
\end{equation}
an equation surely useful if $w$ is constant. Then, in order that the
two sides scale in the same way, it must be $\alpha b = 1$, i.e.
$\rho_c \propto a^{-4}$: $\rho_c$ dilutes faster than $\propto a^{-3}$
because of the leakage of energy onto the $\Phi$ field; that it
dilutes exactly as $a^{-4}$, instead, is a consequence of the ansatz
(\ref{phi1}).  From eq.~(\ref{same}) and the Friedmann equation in the
radiative era
\begin{equation}
{8 \pi \over 3 m_p^2}\,  \rho \, a^2 \tau^2 = 1~,
\label{frie}
\end{equation}
($\rho:$ total density) we then derive
\begin{equation}
{1 \over \beta^2}{\tilde w -1 \over w + 1} = 
{1 \over 2\beta^2}{3w -1 \over w + 1} 
=  {8 \pi \over 3 m_p^2}\,  \rho_c \, a^2 \tau^2
\equiv \Omega_c ~.
\end{equation}
Here $\Omega_c = \rho_c /\rho$ is the (constant) density parameter of
DM (during the radiative era). In order that the ansatz (\ref{phi1})
is allowed, $\Omega_c$ ought to have the value given by this equation.

It is then easy to show that the $\Phi$ field energy density is
\begin{equation}
\rho_d = {\alpha^2 m_p^2 \over a^2 \tau^2 }{1 \over 1+w}~,
\end{equation}
so that its constant density parameter reads
\begin{equation}
\Omega_d = {1 \over 2\beta^2 (1+w)}
\label{omegad}
\end{equation}
showing also that
\begin{equation}
{\Omega_c / \Omega_d} = 3w-1~.
\end{equation}
Quite in general, being
\begin{equation}
w = {\Phi_1/2a^2 - V(\phi) \over \Phi_1/2a^2 + V(\phi) }~,
\end{equation}
$w$ is constant when either the kinetic or the potential term,
(almost) dominate. In the former (latter) case $w=+1 ~(-1)$.

In any coupled DE theory, at high redshift the field is essentially
kinetic and $w=+1$. If uncoupled, its energy density would dilute
$\propto a^{-6}$. Equation (\ref{omegad}) tells us that, instead, in
the presence of coupling, its early density parameter can be constant,
as it dilutes $\propto a^{-4}$.

Altogether, for $w=+1$, we then have:
\begin{equation}
\label{omegs}
\Omega_c = {1 / (2\beta^2) }~,~~~~
\Omega_d = {1 / (4\beta^2) }~,~~~~
(\Omega_c+\Omega_d) \beta^2 = 3/4~.
\end{equation}
In the early Universe the total density parameter is unity. Requiring
then $\Omega_c + \Omega_d < 1$~yields
\begin{equation}
\beta > \sqrt{3}/2 = 0.866~.
\end{equation}
In Paper I we also tested that this solution of eq.~(\ref{c1}) is an
attractor: starting from generic initial conditions, $\Phi$ and
$\rho_c$ rapidly evolve to approach a regime where eqs.~(\ref{omegs})
are satisfied.

We denominate {\it strongly coupled cosmologies} those with $\beta >
0.866$, allowing $ \Omega_c$ and $\Omega_d $ to be constant in the
radiative era. On the contrary, when $\beta^2 < 3/4$ there is no
solution with constant $\Omega_{c,d} :$ the CDM and $\Phi$ field
contributions to the overall density decrease when $a$ tends to zero.

\section{Early expansion end}
The regime described in the previous Section could be present since
ever, e.g. since inflation, and would last forever unless a
non--relativistic component grows to overcome radiative ones.

At variance from Paper I, in this work we shall mostly assume that
this component is warm. Its energy density and pressure read
\begin{equation}
\rho_w = { T_w^4 \over \pi^{2} } \int_0^\infty
dx~x^2 {\sqrt{x^2+(m_w/T_w)^2} \over e^x+1}
~,~~~
p_w = { T_w^4 \over \pi^{2} } \int_0^\infty
dx~{x^4 \over \sqrt{x^2+(m_w/T_w)^2}}{1 \over e^x+1}
\end{equation}
and derelativization occurs when the WDM temperature $T_w$ shifts
below the mass $m_w$ of WDM quanta. Then we gradually achieve the
regime $\rho_w \propto T_w^3$ and WDM density overcomes the radiative
components. A little later, also baryons will do so.

This is the first stage forging the present sharing of densities among
cosmic components. It is also critical to establish when the $\Phi$
field passes from the kinetic to the potential regime. In most
previous work this stage was followed by using an expression of its
self--interaction potential $V(\Phi)$. The potential was often
selected so to allow tracker solutions. 
\begin{figure}
\begin{center}
\includegraphics[scale=0.37]{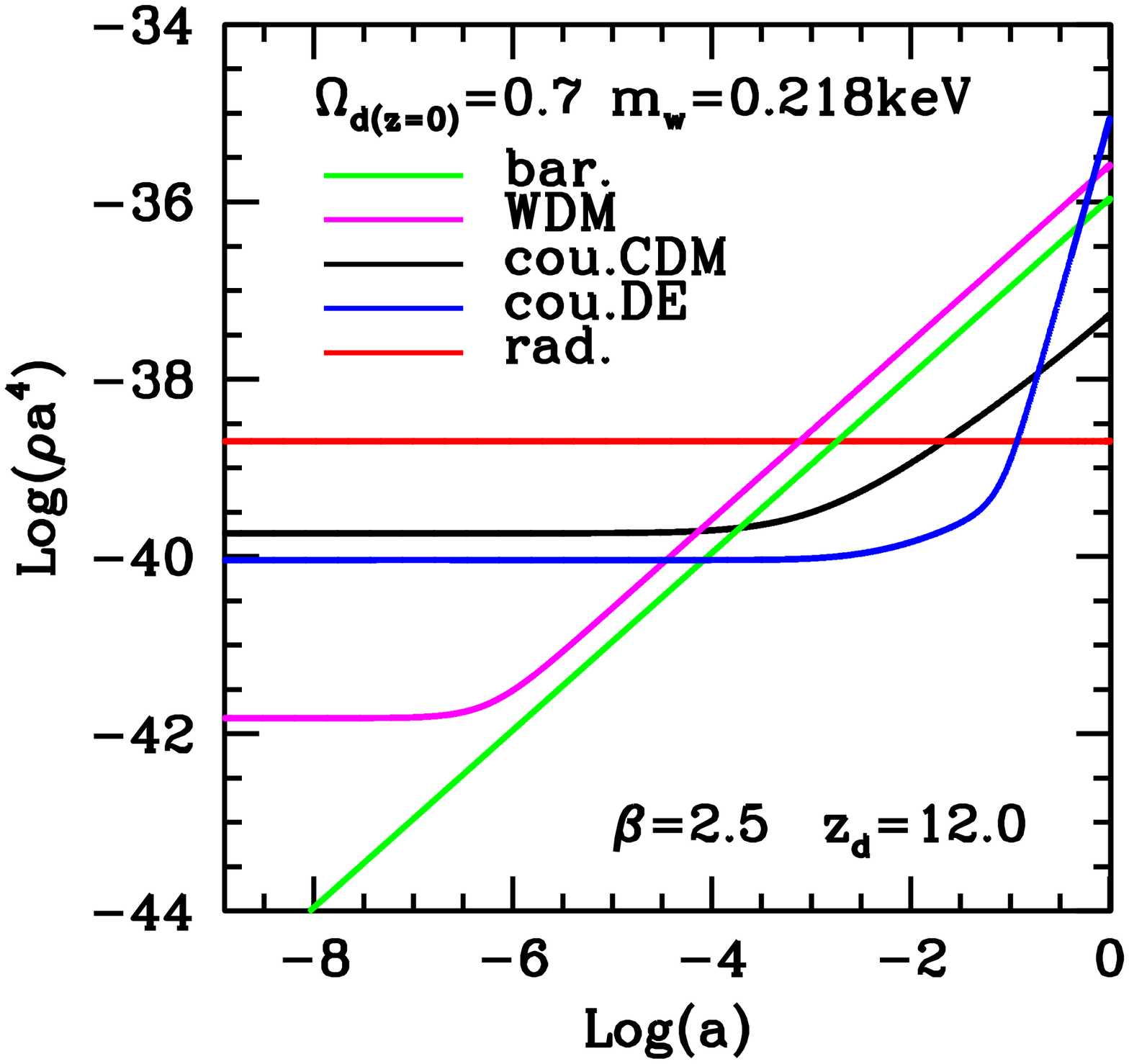}
\includegraphics[scale=0.37]{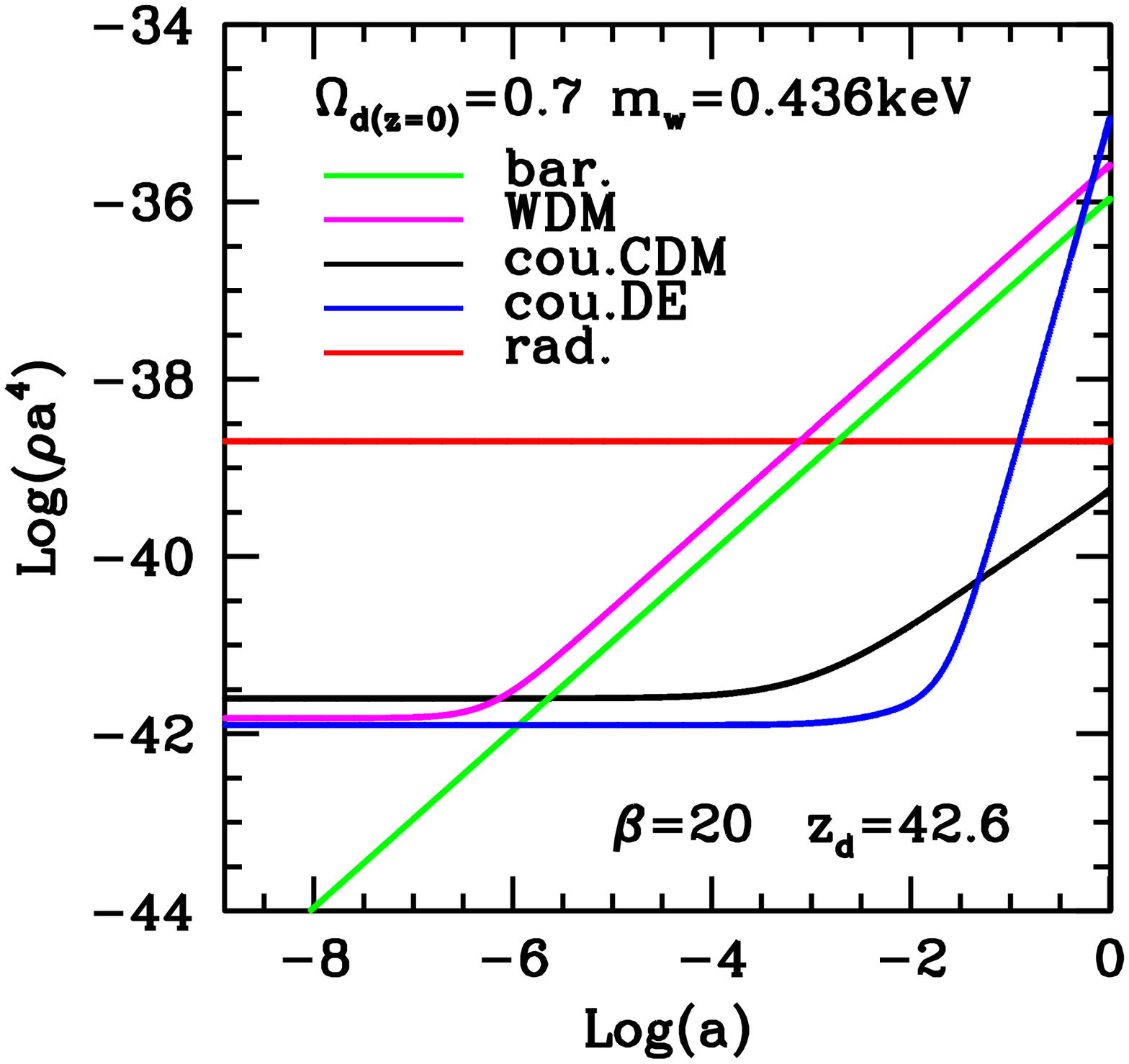}
\end{center}
\vskip -.6truecm
\caption{Evolution of background components in cosmologies with
  coupled CDM and uncoupled WDM (spinor thermal particles with 2 spin
  states). In both cases we took $\epsilon=2.9~$ and, at $z=0$,
  $\Omega_d = 0.7$, $h=0.68$.  Values of $\beta$ and $z_d$ shown in
  the frames; the latter is obtained by suitably tuning the transition
  from kinetic to potential regime of the $\Phi$ field. Densities are
  given in MeV$^4$. For $\beta = 2.5$ the contribution of the coupled
  components to the early density is just below 1 extra massless
  neutrino species. For $\beta = 20$, when WDM is relativistic and DE
  is kinetic, $\Omega_w \simeq \Omega_d,~\Omega_c~.$ }
\label{exe0}
\end{figure}

\begin{figure}
\begin{center}
\includegraphics[scale=0.42]{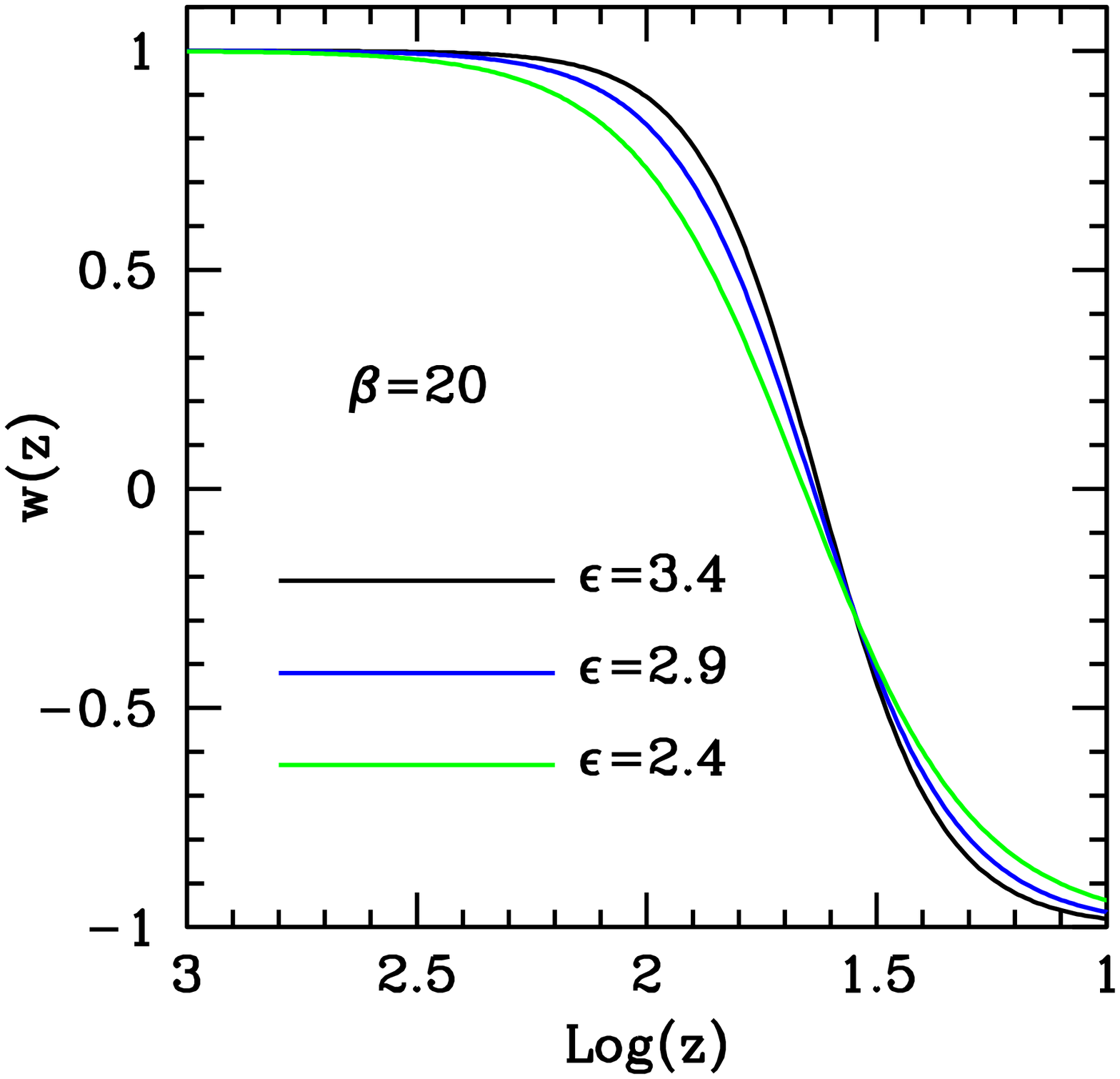}
\includegraphics[scale=0.25]{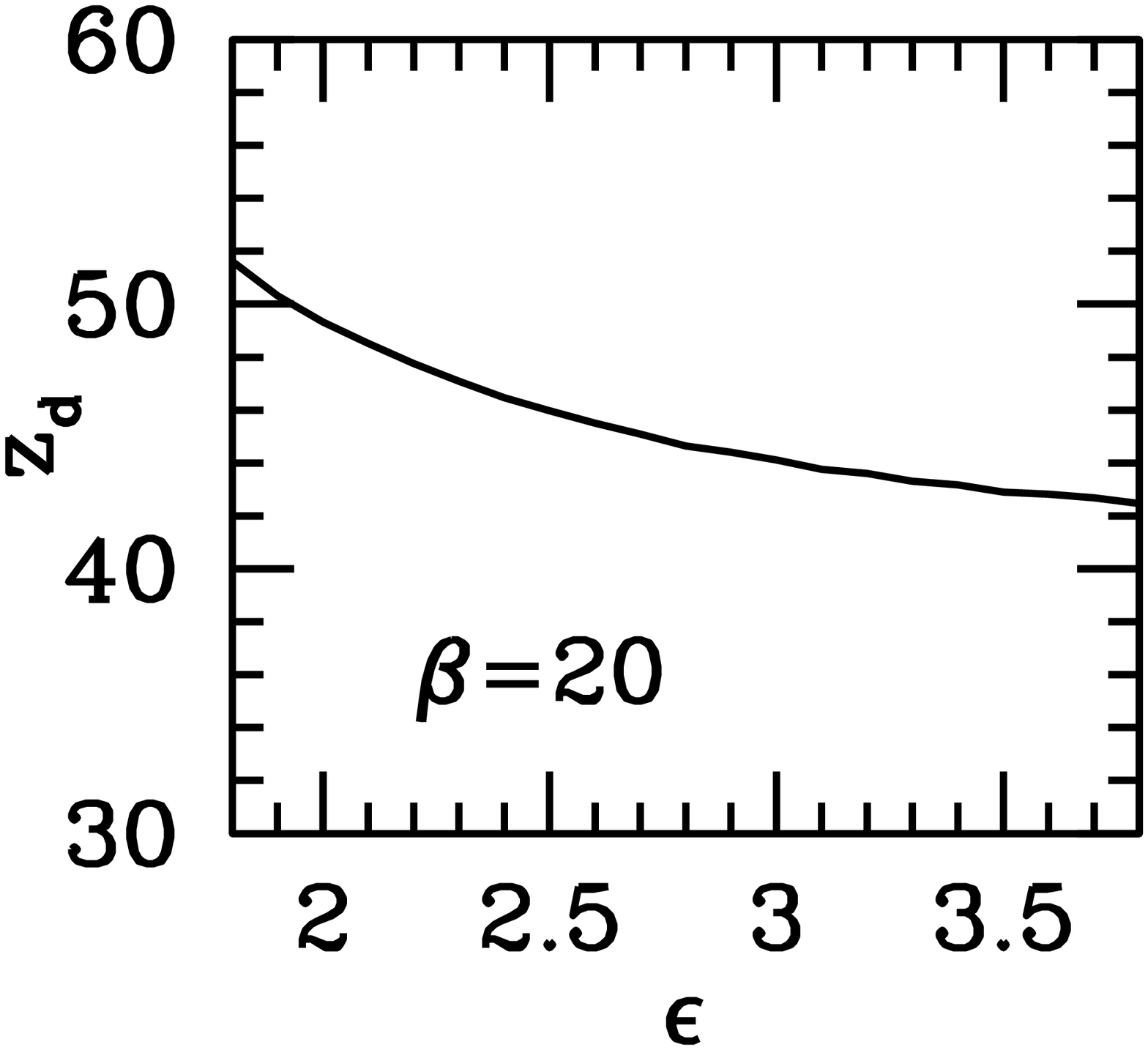}
\end{center}
\vskip -.2truecm
\caption{Transition from $w=+1$ to $w=-1$ of DE state equation. We
  assume $\Omega_d = 0.7$ at $z=0$ and consider the case
  $\beta=20$. At the r.h.s. we show the $w(z)$ dependence for 3
  $\epsilon$ values. At the r.h.s. the values of $z_d$ yielding
  $\Omega_d = 0.7$ are plotted vs.~$\epsilon$.  }
\label{trans}
\end{figure}

Determining the shape of $V(\Phi)$ from observational data is however
(almost) hopeless. We find that the critical feature is rather the
redshift $z_d$ (scale factor $a_d = (1+z_d)^{-1}$)
when the kinetic--potential transition takes place.
Through this paper we shall assume that
\begin{equation}
w = {1-A \over 1+A} ~~~ {\rm with }~ ~ A=\left(a \over a_d\right)^\epsilon
\label{ad}
\end{equation}
so that $w \to +1~(-1)$ for $a \to 0~(a_0=1)$, and $w(a_d) = 0$. The
parameter $\epsilon$, rather than a peculiar shape of $V(\Phi),$ then
fixes the sharpness of the transition. In most of this work we assume
$\epsilon = 2.9~.$ The above expression improves the one used in Paper
I, for having continuous first and second derivatives. With this $w$,
in particular, 
\begin{equation}
\tilde w = {4+(\epsilon -2)A \over 2(1+A)}
\label{Atw}
\end{equation}
The present density of DE is essentially fixed by $a_d$, with a
milder dependence on $\epsilon$. 

In Figure \ref{exe0} we then show the density evolution of all cosmic
components for 2 values of $\beta$: (i) $\beta = 2.5$ is slightly
above the value $\beta = 2.19$ given by
\begin{equation}
\beta^2 = {3 \over 4} \left[2 + {8 \over 7} \left(11 \over 4
  \right)^{4/3} \right] ~,
\end{equation}
which may be considered a phenomenological lower limit, as it yields a
coupled DE--CDM component whose density approaches one extra neutrino
species (the contribution to the background density due to WDM is
disregarded).  (ii) $\beta = 20$, instead, is close to the coupling
strength yielding a coupled DE component with the same early density
of WDM.

Let us notice that WDM derelativization occurs at a redshift $z_{der}
\propto m_w$ so that, if we assume a fixed low--$z$ density parameter
for WDM, the early $\Omega_w \propto m_w^{-1}$, while the early
$\Omega_d \propto \beta^{-2}$. Accordingly, an approximate coincidence
$\Omega_d \sim \Omega_w$ can be mantained only if the WDM particle
mass $m_w \propto \beta^2$.

Figures \ref{trans} then illustrate the dependence of results on the
choice of $\epsilon$, which is quite mild for $\epsilon >\sim 2.5$.
The range of $\epsilon$ can be interpreted as the possible dependence of
results on the shape potential $V(\Phi)~.$

\section{Perturbations}
Let us now consider the evolution of {\it small} perturbations to this
background in a synchronous gauge.  Quite in general, the metric reads
\begin{equation}
ds^2 = a^2(\tau) \left[ d\tau^2 - (\delta_{ij}+h_{ij})dx^idx^j \right]~;
\end{equation}
scalar metric perturbation can then be expanded as follows \cite{Ma}:
\begin{equation}
h_{ij}(\tau,{\bf x}) = \int d^3k~e^{i{\bf k} \cdot {\bf x}} [n_in_j
    h(\tau,{\bf k}) + (n_in_j - \delta_{ij}/3) \, 6\eta(\tau,{\bf k}) ]
\end{equation}
with ${\bf k} = {\bf n}k$. Einstein equations then yield
\begin{equation}
\ddot h + {\dot a \over a} \dot h 
= -{8\pi \over m_p^2}\, a^2 (\delta \rho + 3\delta p)~.
\label{h}
\end{equation} 
The gravity sources to be included in the term $\delta \rho + 3\delta
p$ are: (i) radiation, for which $\delta \rho + 3 \delta p = 2 \rho_r
\delta_r$; (ii) baryons, for which $\delta \rho + 3 \delta p = \rho_b
\delta_b$; (iii) uncoupled CDM or WDM, for which $\delta \rho + 3
\delta p = c_w \rho_w \delta_w$ ($c_w=1$ in the CDM case; in the WDM
case $c_w=2$ until it is ultrarelativistic; derelativization will then
be followed by sharing WDM energy spectrum in a suitable number of
components, chosen to allow Gauss-Laguerre momentum integration); (iv)
coupled CDM, also yielding $\delta \rho + 3 \delta p = \rho_c
\delta_c$; and, finally, (v) the DE field $\Phi$, for which
\begin{equation}
\delta (\rho_\phi+3p_\phi) = 
\delta \left[ 4 {\Phi_1^2 \over 2a^2} - 2V(\Phi) \right] =
4 {\bar \Phi_1  \phi_1 \over a^2} - 2V'(\bar \Phi) \phi~.
\label{phi}
\end{equation}
$\bar \Phi_{(1)}$ being the background field. In principle it is then
$ \Phi_{(1)} = \bar \Phi_{(1)} + \phi_{(1)}$ with $\phi_{(1)}$
accounting for $\Phi$ fluctuations. However, here below, scalar field
fluctuations will be mostly described by the dimensionless variable
$\varphi = (b/m_p) \phi$ and its derivative $\dot \varphi = (b/m_p)
\phi_1$; let us remind that $b=(16 \pi/3)^{1/2}\beta$, according
  to eq.~(1.1). Furthermore the bar in top of $\Phi$ is omitted and the
background field is simply $\Phi_{(1)}$. From eq.~(\ref{h}), for
gravity fluctuations we then obtain
\begin{equation}
\ddot h + {\dot a \over a} \left(\dot h + {6 \over \beta^2} D 
\dot \varphi \right) = -{8\pi \over m_p^2}
  a^2 (2\rho_r\delta_r + \rho_b \delta_b + c_w \rho_w \delta_w + \rho_c
  \delta_c) + \sqrt{ 16 \pi \over 3 } a^2{V'(\Phi) \over m_p \beta}
  \varphi
\label{h1}
\end{equation}
with $D\dot a/a = \Phi_1 b/m_p$ and, when in the kinetic regime, the
last term at the r.h.s. can be simply omitted; otherwise, we can use
the expression (\ref{V'}) for $V'(\Phi)$.

The equations of motion of the cosmic components will then be written
by neglecting massless neutrinos, unessential to understand the
dynamics of the model. Let us soon outline, however, that final
results obtained though a suitable modification of CMBFAST take into
account 3 standard massless neutrino species. The equation of motion
will be written for the case when uncoupled DM is cold, and read:
\begin{equation}
 \dot \delta_r = -{2 \over 3} \dot h - {4 \over 3}kv_r~,~~~ \dot v_r
= {1 \over 4} k\delta_r~,~~~ \dot \delta_w = -{1 \over 2} \dot h~,~~~
\dot \delta_c = -{1 \over 2} \dot h - \dot \varphi -kv_c~,~~~ \dot v_c =
-{\dot a \over a}(1-D) v_c -k \varphi~,
\label{motion1}
\end{equation}
\begin{equation}
\ddot \varphi + 2{\dot a \over a} \dot \varphi + {1 \over 2} \Phi_1
\dot h +k^2 \varphi + a^2 V''_\phi(\Phi) \varphi = 2 \beta^2
\left(\dot a \over a \right)^2 \Omega_c \delta_c~.
\label{motion2}
\end{equation}
In order to obtain equation~(\ref{motion2}) we exploited the fact
that, thanks to the Friedmann equation, $(b/m_p^2)\, a^2 \rho_c = 2
\beta(8\pi/3 m_p^2)\, a^2 \rho\, \Omega_c = 2 \beta^2(\dot a/a)^2
\Omega_c.$ The first two equations (\ref{motion1}) refer to radiation;
here they assume baryons to be tightly bound to photons, an assumption
surely reliable, over most significant $k$ scales, at least up to
matter--radiation equality. From these equations we easily work~out
\begin{equation}
\ddot \delta_r + {1 \over 3} k^2 \delta_r = -{2 \over 3} \ddot h~,
\end{equation}
as is expected, owing to the neutrino neglect.  When gravitation is
negligible, therefore, we expect harmonic oscillations in the
photon--baryon fluid, with period $P=2\pi\sqrt{3}/k$ in respect to
conformal time.

The equation of motion for $\delta_w$ assumes an uncoupled cold
component. On the contrary, WDM fluctuations cannot be described by a
single function $\delta_w$, needing suitable expansions in respect to
particle momenta and spherical harmonics. In the final quantitative
analysis, we shall keep to the standard treatment (see, e.g.,
\cite{BV,Ma}). Let us just notice that, for non--vanishing components,
in the initial conditions $(1/2)\dot h$ shall then be replaced by
$(2/3)\dot h$.

Finally, notice also that, in the kinetic regime, the problematic
$V''$ term can be omitted from the $\varphi$ field equation, on the
last line. We shall return on this point in Section~7.

\section{Out--of--horizon solutions}
Before the entry in the horizon, also the term $k^2 \varphi $, $kv_c$
and $kv_r$ can be disregarded in the field and the former CDM and radiation
equations, while the latter CDM and radiation equations can be disregarded. Let
us then try to solve the system by making the following ansatz:
\begin{equation}
h = A \tau^x~,~~ \delta_r = R \tau^r
~,~~ \delta_w = W \tau^y
~,~~ \delta_c = M \tau^c
~,~~  \varphi= F \tau^f
\end{equation}
From the eqs.~(\ref{motion1}) we obtain
$$
rR\tau^{r-1} = -{2 \over 3} x A \tau^{x-1}
~,~~
yW\tau^{y-1} = -{1 \over 2} x A \tau^{x-1}
~,
$$
\begin{equation}
cM\tau^{c-1} = -{1 \over 2} x A \tau^{x-1} - fF\tau^{f-1}~,
\label{ooh1}
\end{equation}
while the field equation (\ref{motion2}) becomes
\begin{equation}
2f(f+1) F \tau^{f-2} + xA \tau^{x-2} = 4 \beta^2 \Omega_c M \tau^{c-2}
\label{ooh2}~,
\end{equation}
once we replace the background field $\Phi_1 = m_p/b\tau~$. Similarly,
by using again the Friedmann equation, eq.~(\ref{h1}) yields
\begin{equation}
x^2 A \tau^{x-2} + {6 \over \beta^2} fF \tau^{f-2}
+ 6 \Omega_r R \tau^{r-2} + 3 \Omega_w W \tau^{y-2}
+ 3 \Omega_c M \tau^{c-2} = 0~.
\label{ooh3}
\end{equation}
It it easy to see that eqs.~(\ref{ooh1}), (\ref{ooh2}), (\ref{ooh3})
require that
\begin{equation}
x = f = r = y = c 
\end{equation}
eqs.~(\ref{ooh1}), (\ref{ooh2}) also require that
$$
2A + 3R = 0~,~~
A + 2W = 0~,~~
A + 2M + 2F = 0~,
$$
\begin{equation}
xA - 4\beta^2 \Omega_c M + 2x(x+1) F = 0~,
\label{sy1}
\end{equation} 
while eq.~(\ref{ooh3}) yields
\begin{equation}
x^2 A + 6 \Omega_r R 
 + 3 \Omega_w W  + 3 \Omega_c M + (6 x/\beta^2) F = 0~.
\label{sy2}
\end{equation}
The first two eqs.~(\ref{sy1}) allow us to obtain $R$ and $W$ in terms
of $A$. We use them in eq.~(\ref{sy2}), taking also into account
eqs.~(2.15) and assuming $\Omega_w$ to be negligible. The system
(\ref{sy1}), (\ref{sy2}) then yields
$$
A(x^2 - 4 + 3/\beta^2) + F 6x/\beta^2 + M 3/2\beta^2 = 0
$$$$
A x/2 + F x(x+1) - M = 0~,~~~
A + 2F + 2M = 0~,
$$ 
and non--vanishing solutions exist only if we fulfill the dispersion
relation
\begin{equation}
\left|
\matrix{ x^2 - 4 + 3/\beta^2 & 6x/\beta^2 &  3/2\beta^2 \cr
x/2 & x(x+1) & -1 \cr 1 & 2 & 2}
\right| = 0
\end{equation}
easily reordered to obtain
\begin{equation}
(x^2 - 4) [ 2(x^2+x+1) - 3 / 2\beta^2 ] = 0~.
\end{equation}
There are then two cathegories of out--of--horizon fluctuation modes:

\noindent
modes (a): $x = \pm 2$, being $\beta$ independent;

\noindent
modes (b): $x = (1/2)[-1 \pm 3^{1/2} (1/\beta^2 -1)^{1/2}]$~.

In the case of a cold uncoupled DM component, for all modes it is
\begin{equation}
R = -(2/3) A~,~~ W = (-1/2) A~,
\label{any}
\end{equation}
also implying $R = (4/3) W\, $, as expected. If, instead, the
uncoupled DM component is warm and still relativistic when I.C. are
built, it shall be
\begin{equation}
W = R = -(2/3)A~.
\label{any1}
\end{equation}
In the case of the increasing (a) mode, we have
\begin{equation}
\label{m2}
M = -(3/14) A~,~~~ F = -(2/7) A~,
\end{equation}
while, by comparing these equations with eqs.~(\ref{any}) or
(\ref{any1}) we see that
\begin{equation}
M = (3/7) W
\end{equation}
(in the case of relativistic WDM, the coefficient 3/7 should be
replaced by 4/7). Coupled--CDM fluctuations, therefore, are
approximately half of uncoupled--DM. In coupled DE models with
$\beta<\sqrt{3}/2$, coupled--CDM fluctuations are enhanced, in respect
to uncoupled components, by a $\beta$ dependent factor. Here we find
an opposite, $\beta$--independent behavior.

The (a) mode with $x=-2$ is clearly decreasing. (b) modes yield a real
$x$ only in the interval $\sqrt{3}/2 < \beta < 1$. In this interval,
the greatest possible $x$ is 0 and is found at the limit $\beta =
\sqrt{3}/2$. On the contrary, for $\beta > 1$ we find complex $x$
values, which can be combined to yield
\begin{equation}
(\tau/\tau_i)^x = (\tau/\tau_i)^{-1/2} \left\{ A
  \cos[Q\ln(\tau/\tau_i)] + B \sin[Q\ln(\tau/\tau_i)] \right\} ~~{\rm
    with}~~~ Q = [3(1-1/\beta^2)]^{1/2}~,
\end{equation}
$A,~B$ being arbitrary constants and $\tau_i$ a reference time.  They
are however decreasing solutions, comprising an oscillatory behavior
whose physical meaning is hard to realize.

Therefore, out of horizon initial conditions can be set by assuming a
growth $\propto \tau^2$, with fluctuation amplitudes ruled by
eqs.~(\ref{any}),(\ref{m2}), and independent from
$\beta~~(>\sqrt{3}/2)$ value. This is quite alike standard models in
the radiation dominated expansion stages.

\section{Pre-- and post--recombination evolution: a semi--qualitative approach}
Initial conditions can then be applied to the system
(\ref{h1})--(\ref{motion2}).  It is then convenient to define the
variables
\begin{equation}
K(\tau) = 2\beta^2 \Omega_c~~~{\rm and} ~~~ D(\tau) {\dot a \over a} =
\Phi_1 {b \over m_p}
\end{equation}
which, while the initial (pseudo--)stationary regime persists, are $K
= D = 1~.$

By using them and excluding the terms containing the potential $V$,
eqs.~(\ref{h1}) and (\ref{motion2}) read
\begin{equation}
\label{DK}
\ddot h + {\dot a \over a} \left( \dot h + {6 \over \beta^2} 
D \dot \varphi \right) + \left({\dot a \over a} \right)^2
\left[ \left(1 - \Omega_w - {K \over 2 \beta^2} \right)6 \delta_r 
+ 3\, \Omega_w c_w \delta_w + 3\, {K \over 2 \beta^2} \delta_c \right] = 0~,
\end{equation}
\begin{equation}
\ddot \varphi + 2 {\dot a \over a} \left( \dot \varphi + {D \over 4}
\dot h \right) + k^2 \varphi 
= \left( \dot a \over a \right)^2 K \delta_c~.
\end{equation}
Here, the coefficient $c_w$ depends on the state equation of uncoupled
DM.  The radiation and uncoupled DM eqs. are unchanged.

These equations 
\begin{figure}
\begin{center}
\vskip -.4truecm
\includegraphics[scale=0.365]{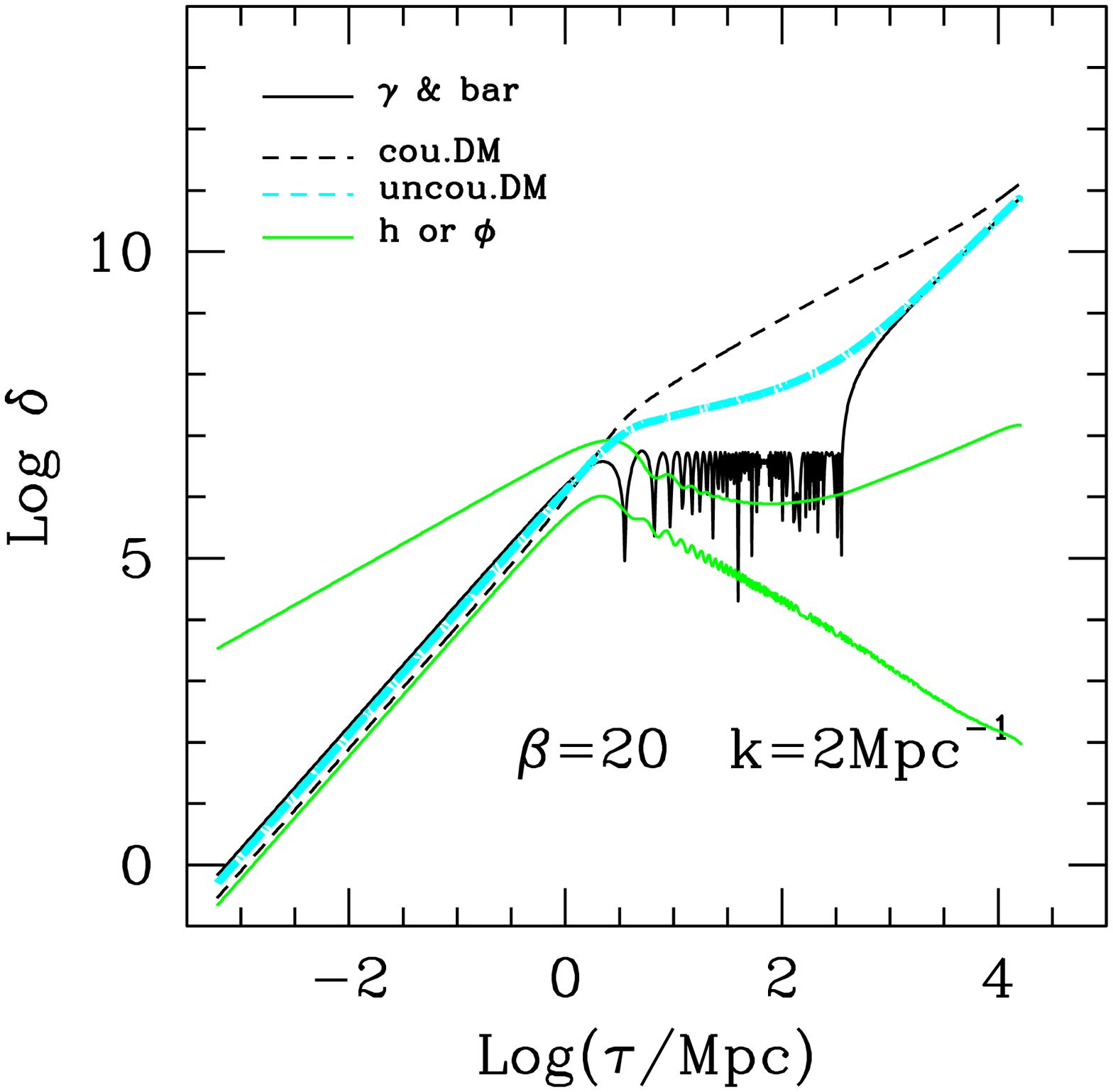}
\includegraphics[scale=0.365]{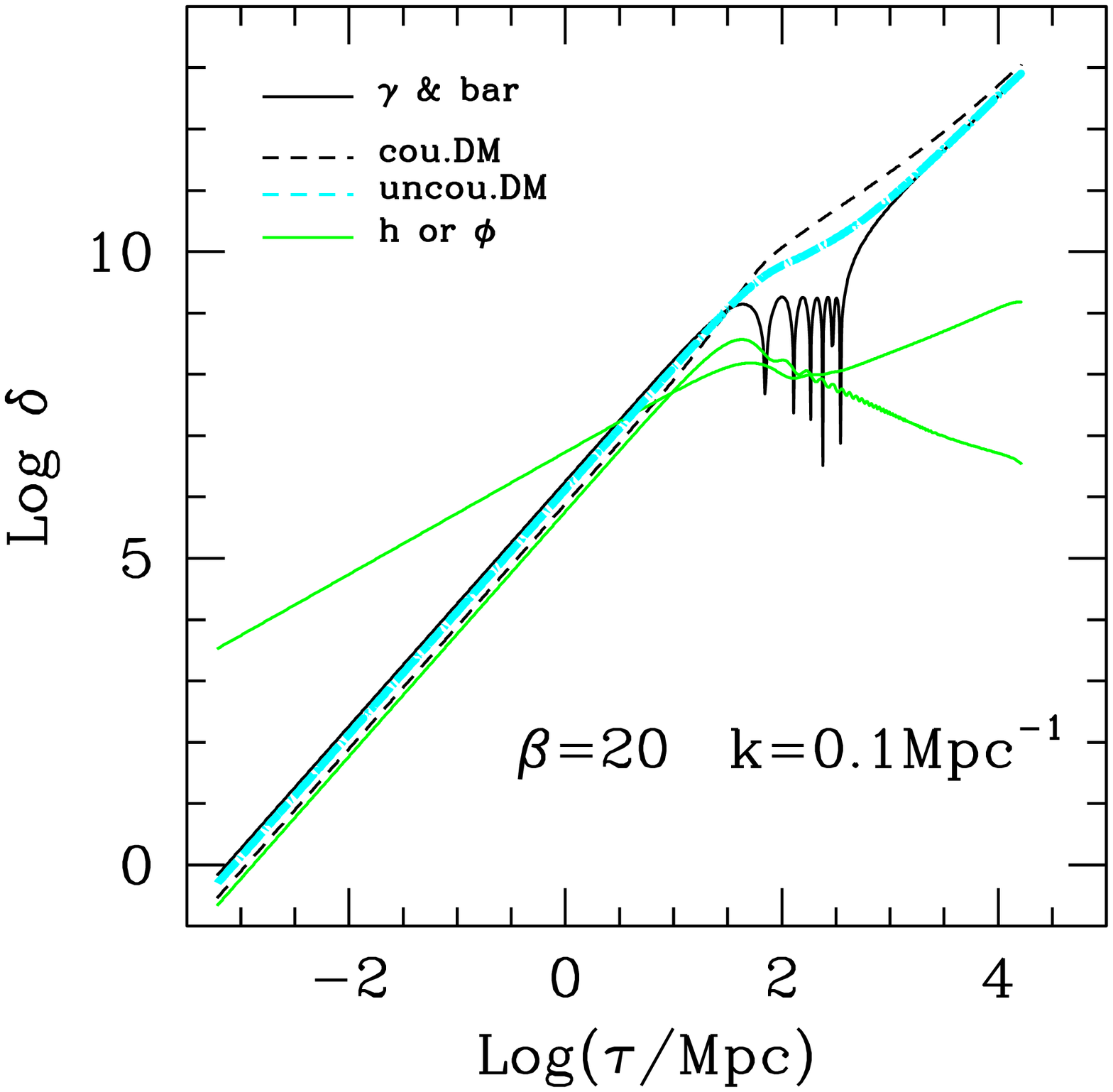}
\includegraphics[scale=0.365]{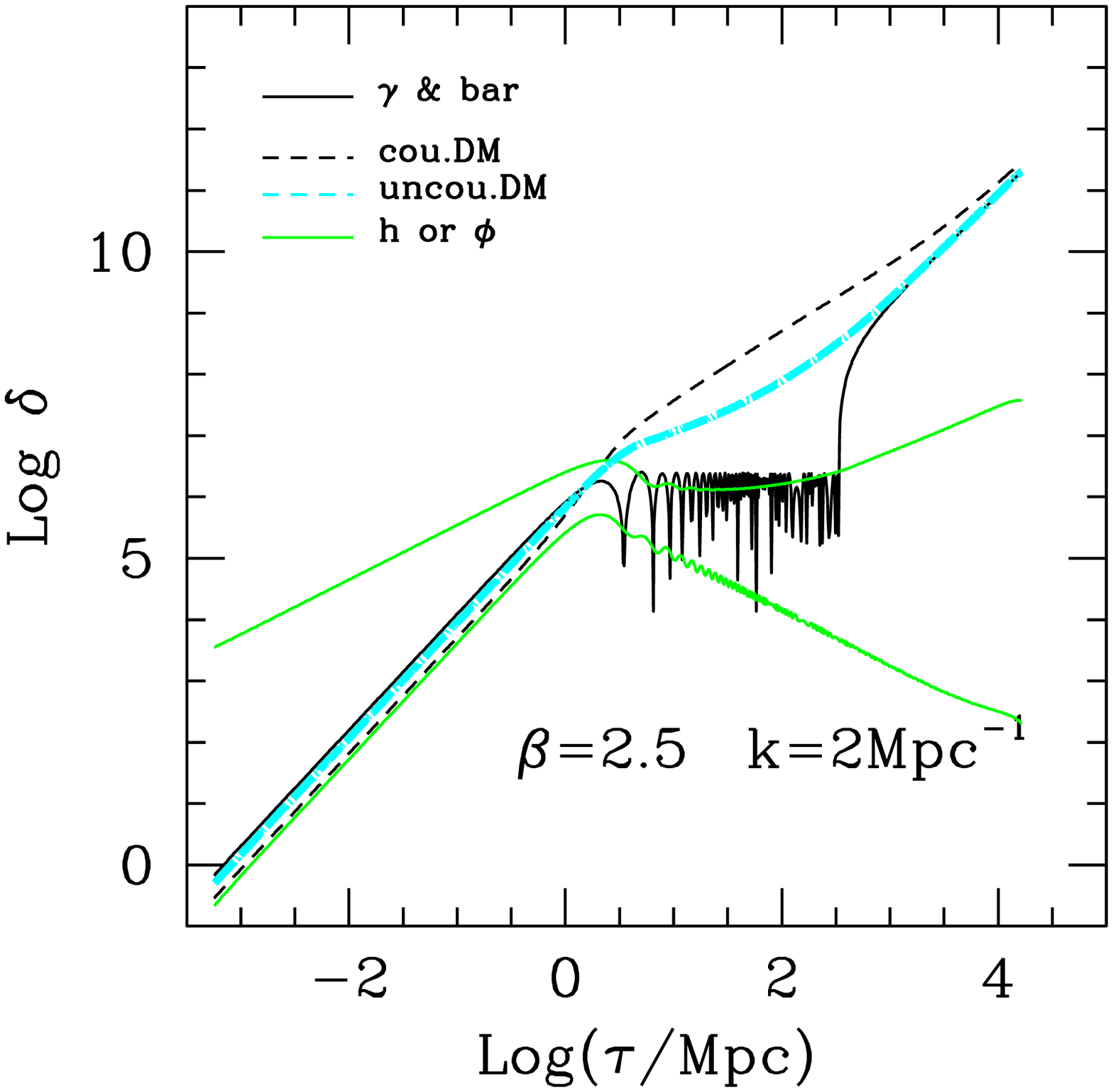}
\includegraphics[scale=0.365]{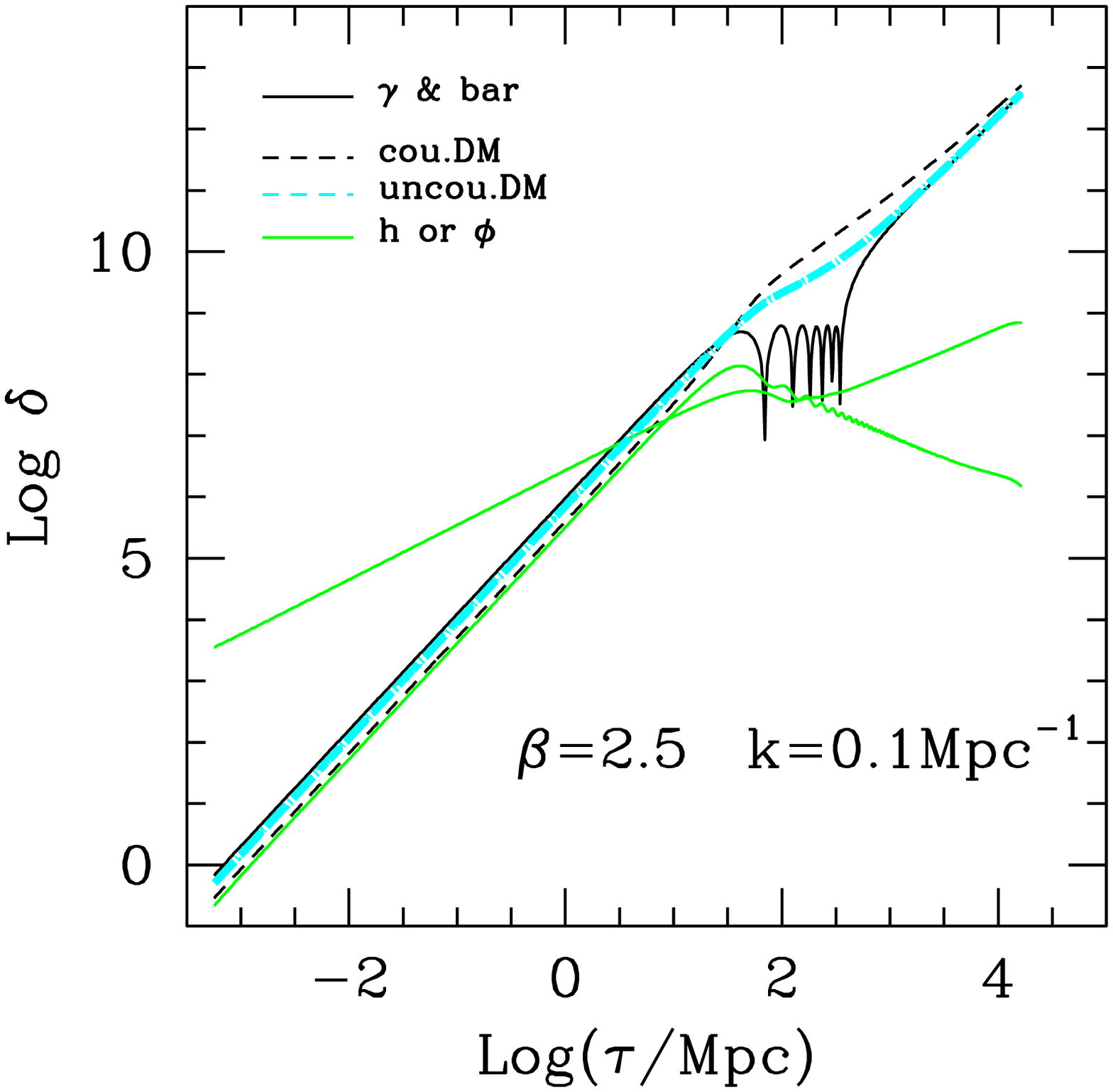}
\end{center}
\vskip -.4truecm
\caption{Fluctuation evolution when uncoupled DM is cold; the moduli
  $|\delta|$, $|\dot h|$, and $|\varphi|$ are plotted for 2 values of
  $\beta$ and two scales $k$. All plots show that coupled DM exhibits
  almost no Meszaros' effect: its fluctuation continue to grow, quite
  rapidly, after the entry in the horizon, also when radiation density
  widely exceeds its density. Such growth has indirect effects also on
  uncoupled DM evolution, which feel the increasing gravity of coupled
  DM. Indirect effects however weaken for greater $\beta$, as coupled
  DM density $\propto \beta^{-2}$. In all plots the present $\Omega_d=
  0.7$.}
\label{lowzevo}
\vskip -.3truecm
\end{figure}
allow us to study fluctuation evolution at any $z$, also when $D,~K
\neq 1$, until the kinetic--potential transition of DE. They allow
also for a component breaking the inizial (pseudo--)stationarity,
provided we assume it to be cold or that we deal with a scale $k$
where WDM free streaming is absent.

We shall now briefly discuss the physical behavior of radiation,
baryons, uncoupled and coupled DM, DE and gravity, by using an 11
component system of linear differential equations whose variables
are:
$$
{\rm for~ the~ background:}~~~a,~\rho_c,~\Phi_1 \equiv \dot \Phi~;
~~~~~~~~~
$$
\vskip -.8truecm
$$ {\rm for ~density~ fluctuations:}~~
\delta_r,~v_r,~\delta_w,~\delta_c,~v_c,~\varphi,~\dot \varphi,~\dot h
$$ The radiation--baryon component will be treated as a fluid with
state parameter $w_R \equiv 1/3$ until $z = 1100$. Afterwards, we
neglect radiation, assuming that fully decoupled baryon fluctuations
$\delta_b = (3/4) \delta_r$ obey a pressureless equation. This option
inhibits predictions on CMB fluctuations, but allows us a few tests on
the pre-- and post--recombination evolution, described in the next
Section.

To go beyond this approximate post--recombination treatment, as well
as in the quantitative treatment, we need to reintroduce the potential
terms and, namely, an expression to replace the term containing
$V''(\Phi)$ when we focus on the $w(a)$ behavior (approximated by the
expression (\ref{ad})), tentatively disregarding the hardly detectable
shape of $V(\Phi).$

The final part of this Section will be devoted to elaborating this
expression.  Let us first comment on the results of the simplified
dynamical system as shown in Figure \ref{lowzevo}.

All plots show that coupled--DM exhibits almost no Meszaros' effect
\cite{meszaroseff}: on all mass scales below that entering the horizon
at matter--radiation equality, fluctuation amplitudes are almost
frozen until matter exceeds radiation density. Let us remind that
Meszaros' effect is critical in shaping the transfer functions.

As a matter of fact, radiation fluctuations, after entering the
horizon, turn into sonic waves, so that their average amplitude
vanishes. When the dominant cosmic component is no gravity source,
other components can cause just a modest push.

The freezing period is unavoidibly longer for smaller scales, spending
more time below the horizon scale before matter--radiation equality,
so that the transfer function decreases at increasing $k$ values. Of
course, in top of this basic ingredient a number of other effects play
a suitable role. Baryons, neutrinos or other specific component add
specific details on the above basic structure.

The reason why coupled--DM fluctuation continue to grow, quite
rapidly, after the entry in the horizon, is visible in the 4--th
eq.~(\ref{motion1}). The gravitational push set by $\dot h/2$ is
there increased by $\dot \phi$, i.e., there is an additional 
force acting just between CDM particles.
Accordingly, coupled CDM fluctuations are a sufficient source to cause
their own growth.

Such growth has indirect effects also on uncoupled DM and baryon
evolution: CDM particles act on them just through ordinary
  gravity, but both of them feel the increasing gravity of 
  wider coupled DM fluctuations. If we compare different plots,
we however see that indirect effects however weaken for greater
$\beta$, as coupled the DM density $\rho_c \propto \beta^{-2}$.

Before concluding this Section, let us focus on the expression to be
used in place of $V''$, when the potential is unknown, but $w(a)$ is
given. Let us then consider eq.~(\ref{c1}), derived in Paper I, in
association with the equation used there to eliminate the $V'$ term
from it. The two equations, reading
\begin{equation}
-a^2 V' = \ddot \Phi + 2 {\dot a \over a} \dot \Phi - C\rho_c a^2~,
\end{equation}
\begin{equation}
{1 + w \over 1 - w} a^2 V' = \ddot \Phi - {\dot a \over
  a} \dot \Phi - {1 \over 1-w^2} {dw \over da} \dot a \dot \Phi~,
\end{equation}
can be subtracted to obtain an expression of $V'$ not incluting $\ddot
\Phi$ (this will prevent the need of considering triple derivatives of
$\Phi$). In this way we obtain
\begin{equation}
\label{2v'}
2 V' = -\left[{ a \over (1+w)} {dw \over da} +3(1-w)\right] {\dot a
  \over a^3} \dot \Phi +(1-w) C\rho_c~.
\end{equation}
$V'$ can then to be derived in respect to $\tau$ and divided by $\dot
\Phi,$ so obtaining $V''(\Phi).$ The general expression is however
cumbersome and useless; it is rather convenient to replace soon the
expression (\ref{ad}) in eq.~(\ref{2v'}). This yields
\begin{equation}
2V' = {A \over 1+A} \left[ \epsilon_6 {\dot a \over a^3} 
\dot \Phi+2C \rho_c \right]
\end{equation}
with $\epsilon_6 = \epsilon-6$.  It will then be 
\begin{equation}
2V'' = {A \over 1+A}
\left\{ 
{\dot a \over a} {\epsilon \over 1+A} 
\left[\epsilon_6 {\dot a \over a^3} 
+ 2C {\rho_c \over \dot \Phi} \right]
+ \left[  {\dot a \over a^3} {\ddot \Phi \over \dot \Phi}
+ {d \over d\tau}  ({\dot a \over a^3}) \right] \epsilon_6
+ 2C {\dot \rho_c \over \dot \Phi} \right\}
\end{equation}
and this equation is soon suitable to numerical evaluations; in fact,
it contains variables any numerical algorithm however needs to
evaluate, apart of
\begin{equation}
 {d \over d\tau}  ({\dot a \over a^3}) = -{4 \pi \over 3 m_p^2}
(5 \rho + 3 p)
\end{equation}
which requires an explicit expression of the pressures $p$ of
all cosmic components.

We shall however further discuss this point after giving some
numerical results. In Figure \ref{lowzevo} we show the expected
evolution of fluctuations for all components, under the above
assumptions. Here also a case with a low $\beta$ value is considered.
Low $\beta$'s weaken the Meszaros effect and cause a smaller
binding of the transfered spectra.

\section{Quantitative results}
The simplified 11--eqs. treatment is effective to understand the basic
physical effects, but unsuitable to predicting CMB fluctuation
spectra, treating the case when the uncoupled DM component is warm,
and to allow us a detailed comparison with other models.

In order to achieve such aims, we suitably corrected the public
program CMBFAST. The program allows for both massless and massive
neutrinos, and the latter facility can soon be used to follow WDM
fluctuations; CDM and DE equations are however to be widely modified,
starting from initial conditions, both for fluctuations and background
components.

\begin{figure}
\begin{center}
\vskip -.4truecm
\includegraphics[scale=0.65]{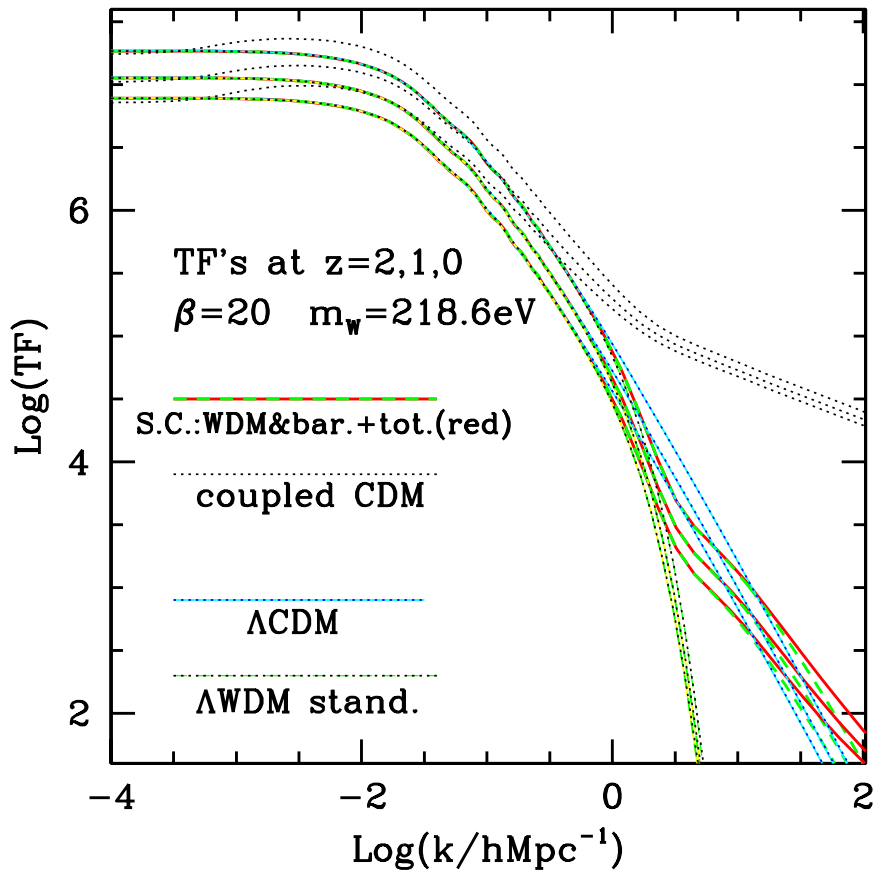}
\end{center}
\vskip -.4truecm
\caption{Transfer function for a strongly coupled cosmology with
  $\beta=20$ and $m_w=218.6$ compared with the transfer functions of
  $\Lambda$CDM and $\Lambda$WDM models with identical cosmic
  parameters; in particular:
  $\Omega_{0b}=0.05,~\Omega_{0\Lambda}=0.7,~n_s=0.96,~H=70.9\, $(km/s)/Mpc.}
\label{TF20}
\vskip -.3truecm
\end{figure}
\begin{figure}
\begin{center}
\vskip -.4truecm
\includegraphics[scale=0.65]{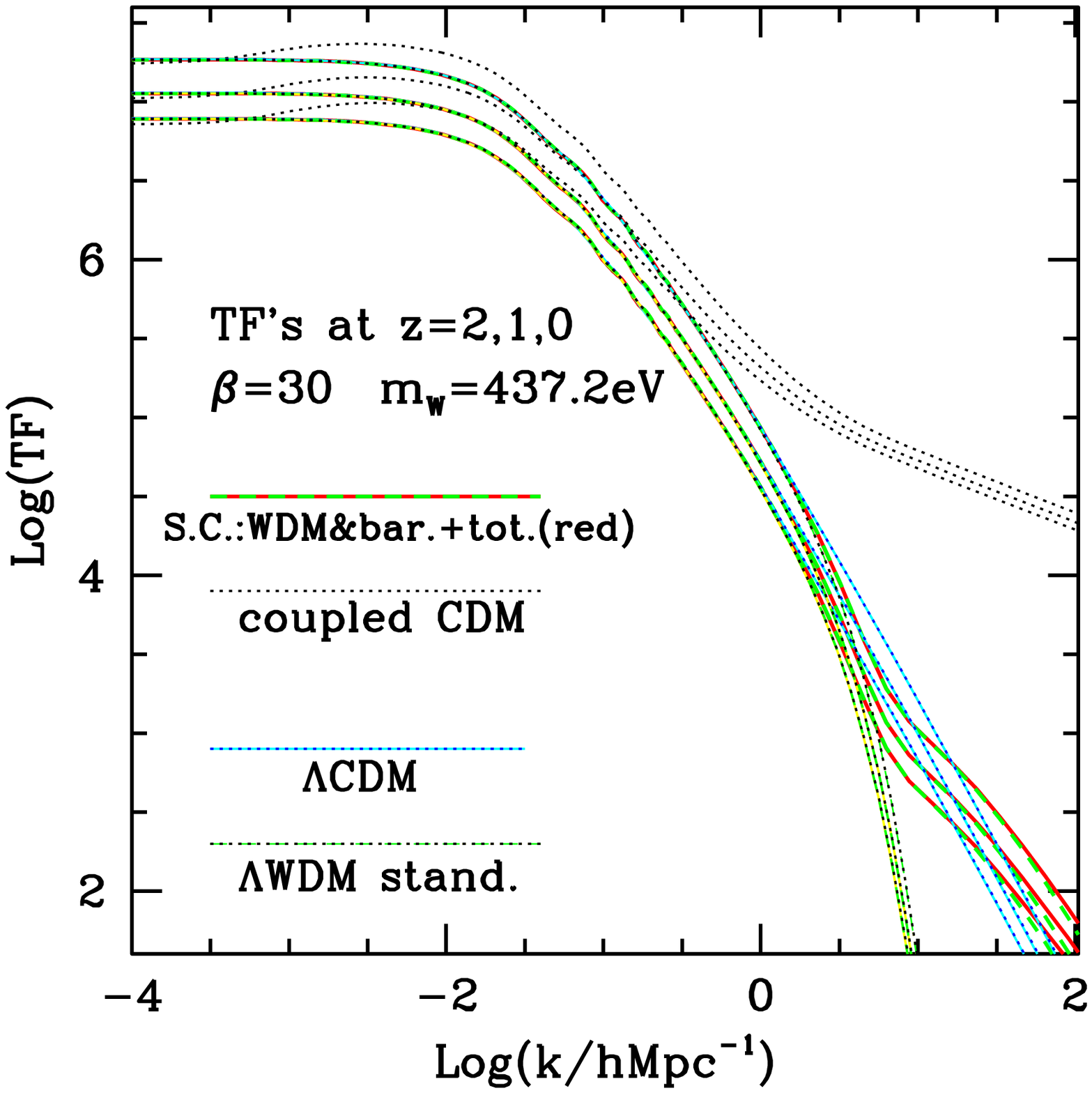}
\end{center}
\vskip -.4truecm
\caption{As the previous Figure, for $\beta=30$ and $m_w=437.2$.
  Notice the (slight) shift on the scale where the S.C. cosmology
  deviates from $\Lambda$CDM. }
\label{TF30}
\vskip -.3truecm
\end{figure}
In Figure \ref{TF20} and \ref{TF30} we give two examples of transfer
functions. In both of them the uncoupled DM is warm, being made of
particles with mass $m_w = 218.6$ or 437.2~eV (corresponding to
$g*=400$ or 800 for 2 spin states; $g*$ is the assumed number of
effective spin states at WDM hot decoupling), respectively. The
coupling is also different, being $\beta = 20$ and 30,
respectively. The simultanous shift of $m_w$ and $\beta$ aims to
provide similar early density parameters for WDM and the coupled
components, in both cases. There is no cogent reason to follow such
prescription, which just alludes to a possible correlated origin for
all dark components, however characterized by different spin and
different numbers of spin states.

The transfer functions of strongly coupled cosmologies are widely
different for the coupled CDM and the other components (WDM and
baryons). The difference starts at a scale $k \sim 10^{-3}$, the scale
entering the horizon at matter--radiation equality. For any greater
$k$, coupled CDM fluctuations, growing between horizon entry and
equality, gradually become greater and greater than WDM and baryons.
The excess amplitude, however, is somehow moderated by the fact that
WDM fluctuations do not grow significantly and, at variance from
radiation, are also gravity sources. When we approach a scale, where
WDM freely streams, the coupled CDM growth has a further burst.

In both plots, coupled cosmologies are compared with a standard
$\Lambda$CDM cosmology and with a $\Lambda$WDM cosmology with the same
parameters. The latter cosmologies differ from $\Lambda$CDM above a
suitable $k$, corresponding to the scale where WDM start to undergo a
free streaming process. 

In strongly coupled cosmologies, the free streaming suppression is
soon balanced by the effect of coupled CDM fluctuation gravity. As
soon as WDM particles become non--relativistic, they re--fall in the
potential wells created by CDM. This is similar to baryons falling in
CDM potential wells after decoupling from radiation, in $\Lambda$CDM
models. The main difference is that, in these latter cosmologies, soon
after decoupling, CDM is the dominant component by far. On the
contrary, in S.C. cosmologies, the coupled CDM component is just a
minimal part of the cosmic substance. Therefore, WDM and baryon
fluctuations, although becoming greater than in $\Lambda$CDM, never
attain the CDM fluctuation level.

Notice also that, up to $k \sim 10\, h$Mpc$^{-1}$, the discrepancy
between the total transfer function and the transfer function for WDM
and baryons is negligible. At greater $k$ values, however, although
CDM is just a few permils of the cosmic materials, its fluctuations
are so wider, to cause a discrepancy between baryons--WDM and total
transfer functions.
\begin{figure}
\begin{center}
\vskip -.4truecm
\includegraphics[scale=0.47]{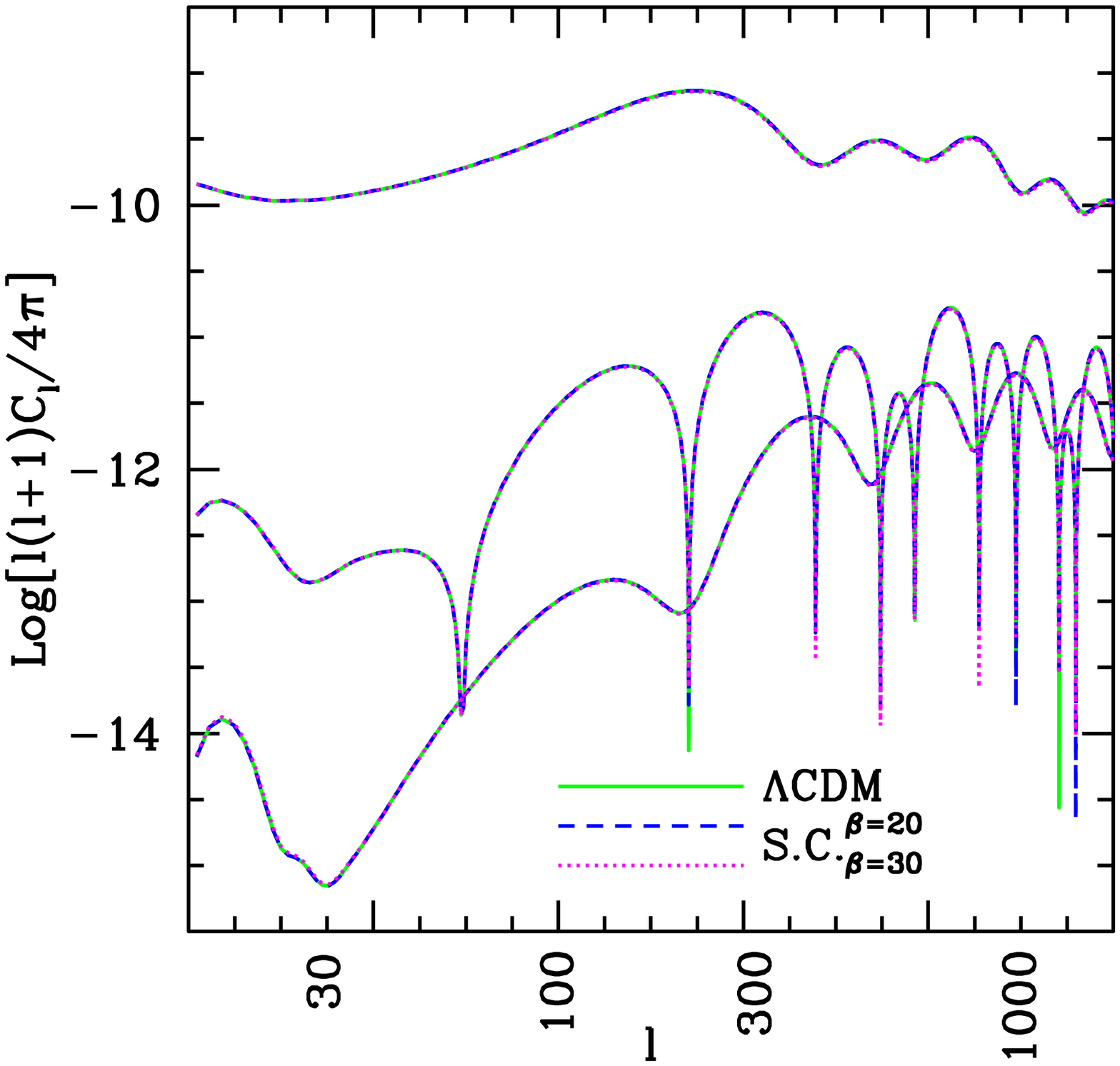}
\includegraphics[scale=0.47]{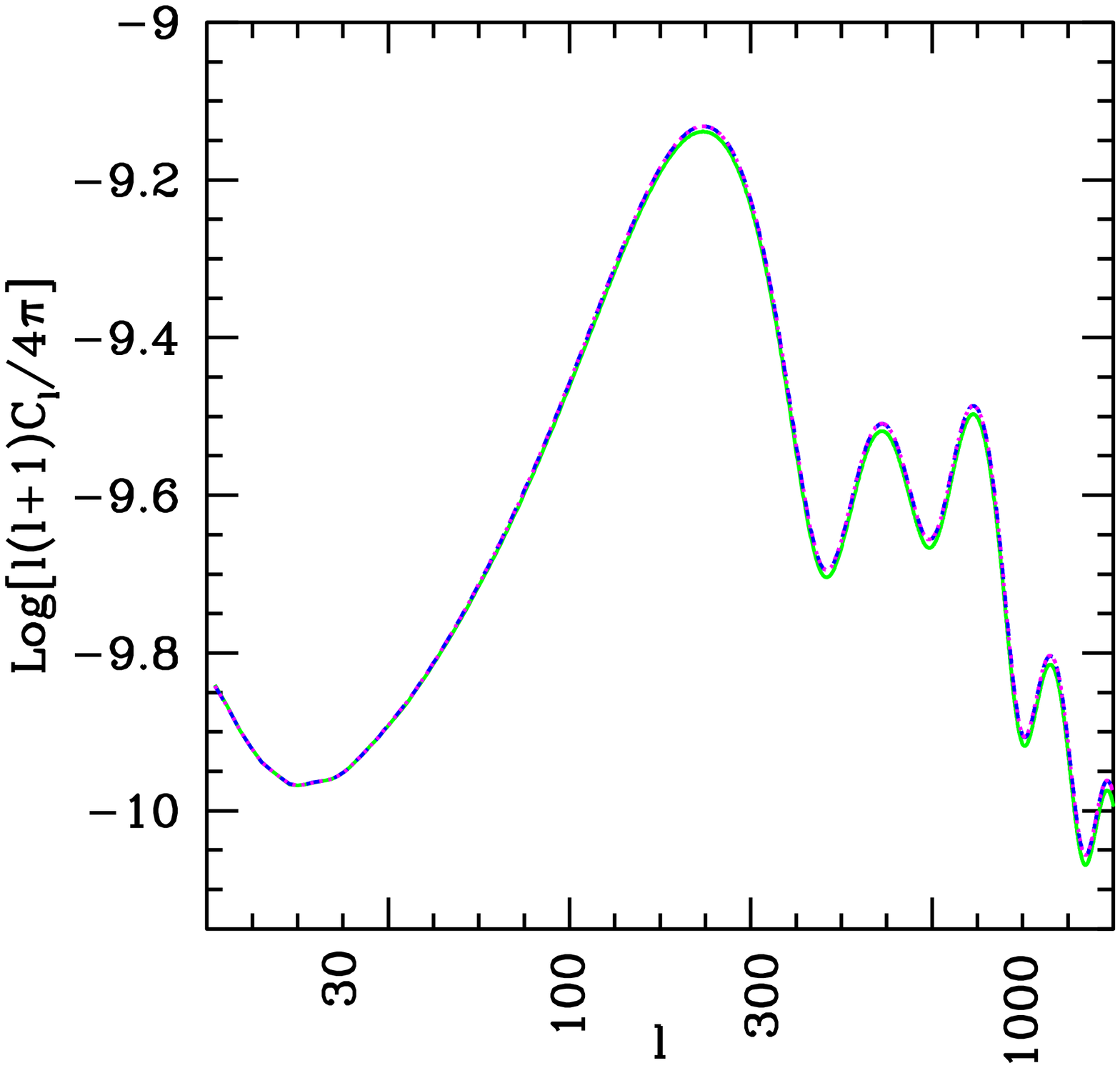}
\end{center}
\vskip -.4truecm
\caption{$C_l$ angular spectra compared. Color selection in the bottom
  plot as in the top one. The only tenuous discrepancies between
  S.C. cosmologies and $\Lambda$CDM are visible in the lower plot,
  where the $EE$ spectrum is expanded. They could be easily
  compensated by a slight change in $n_s$ or the optical depth
  $\tau~.$ }
\label{CL}
\vskip -.3truecm
\end{figure}

Let us however soon outline that the large $\delta_c$ values do not
correspond to a large overall density fluctuation $\delta\rho_c =
\rho_c\delta_c $. In fact, although the ratio $\delta_c/\delta_w$ 
  nearly increases $\propto \beta^2$, the density $\rho_c$ is almost
proportional to the early density parameter $\Omega_c \propto
\beta^{-2}$. Altogether, therefore, the density fluctuations
$\delta\rho_c $ are just slightly decreasing with $\beta$.

A comparison between Figures \ref{TF20} and \ref{TF30} shows a similar
trend in the modification of the transfer function in respect to
$\Lambda$CDM. At a scale fixed by $m_w$ the S.C. transfer function
becomes smaller than $\Lambda$CDM, with a deficit never exceeding one
order of magnitude. This deficit is however localized and, at greater
$k$'s, the transfer function regains and overcomes the $\Lambda$CDM
level. This behavior has been found to be typical off all
S.C. cosmologies where early $\Omega_c$ and $\Omega_w$ are similar. By
increasing $\beta$ (and consequently $m_w$), the feature gradually
displaces towards greater $k$ values.

In Figure \ref{CL} we then compare the CMB angular spectra $C_\ell$
among the same S.C. cosmologies of the previous two Figures and
$\Lambda$CDM. The upper Figure shows $C^{TT}_\ell$, $C^{TE}_\ell$, and
$C^{EE}_\ell$, from top to bottom. The lower Figure, instead, just
concerns $C^{TT}_\ell$.

In the upper Figure the models are barely undistinguishable. Some
slight discrepancy is appreciable in the lower plot, thanks to the
magnification in the ordinate scale. Such minimal discrepancy would be
easily compensated by a tiny shift in the primeval spectral index
$n_s$ or in the cosmic opacity, here assumed to be $\tau =0.089$
anywhere. We may guess that such --almost unperceivably-- greater
$C_\ell$ amplitude in S.C. cosmologies reflects an increase of
radiation fluctuation amplitude due to the large fluctuations in
coupled--CDM.

Altogether we may conclude that, while the transfer function
peculiarities are significant and deserve further discussion, there is
no significant change for CMB spectra in respect, e.g., to
$\Lambda$CDM.

\section{Discussion}
Let us compare the cosmological picture decribed here with more
standard scenarios, by distinguish between (a) conceptual issues and
(b) data fittings.

Let us start from the issue which could appear more controversial and
contrived: the presence of two DM components. The presence of multiple
DM components, however, has been recently advocated by several authors
(see, e.g., \cite{brookbaldi,Maccio':2012uh,blennow}), for precise
observational reasons on which we shall focus when coming to the (b)
point. Our whole approach to the dark cosmic components, however,
avoids any reference to hardly measurable entities; e.g., no specific
self--interacting potential for the $\Phi$ field is assumed,
preferring to refer to its state equation $w(a)$, surely closer to
observations. In a similar way, even if the dual DE--CDM component
could be a single substance (e.g., modulus and phase of a complex
scalar field \cite{axions}), here we keep on the phenomenological side
and treat it as 2 separate components, with a constant coupling
$\beta$ allowing them a suitable energy exchange.

Quite a few advantages of this approach, in respect to more standard
ones, seem however clear. First of all, almost no component needed to
describe today's phenomenology is peculiar of our epoch. This is
surely true for DE, but also for CDM and WDM, the only exception being
baryons. Moreover, radiation, neutrinos, DE, CDM and WDM, until a
fairly recent epoch, kept fixed proportions: the scale factor increase
diluted them all $\propto a^{-4}$. Photons and neutrinos are surely
the dominant components in this (pseu\-do--)sta\-tionary expansion,
but the reason is clear: their densities were enriched by the heating
due to heavier particle decays. This agrees with our choice to take
close values for DE, CDM and WDM early densities, just assuming a
suppression factor for WDM due to its early (hot) decoupling.
Accordingly, {\sl this approach allows for a simultaneous --and, therefore,
possibly correlated-- origin for all dark components.}

\begin{figure}
\begin{center}
\vskip -.4truecm
\includegraphics[scale=0.6]{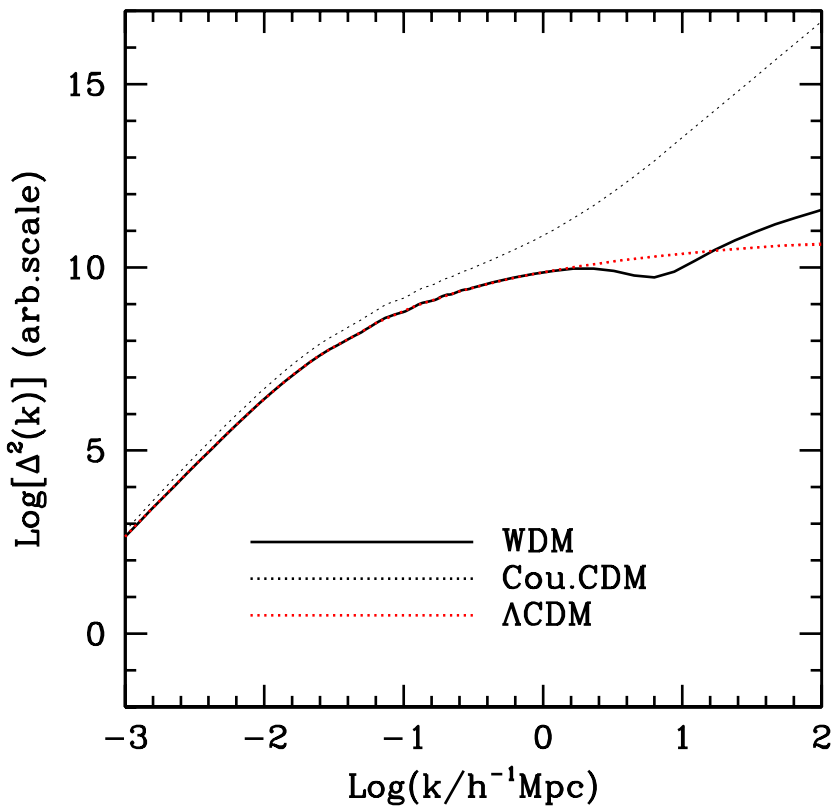}
\end{center}
\vskip -.4truecm
\caption{$\Delta^2(k)$ spectral functions in S.C. cosmologies and
  $\Lambda$CDM.  }
\label{D2}
\vskip -.3truecm
\end{figure}
The break of the early (pseudo--)stationary expansion apparently
requires some tuning, to meet observational features: the
kinetic--potential transition of DE, the rise of baryons, and WDM
derelativization must have followed a precise order. Once again,
baryons are a problem; not worse than in any other cosmological
framework, however. On the contrary, when we come to transitions
concerning DE and WDM, their quasi--coincidence does not appear so
awkward. If 
\begin{equation}
\label{pot}
V(\Phi) \simeq m^2 \Phi^2~, 
\end{equation}
or a similar term is part of the self--interaction potential, we meet
the transition when the decrease of the kinetic energy density $\dot
\Phi^2/2a^2 \propto a^{-4}$ lends relevance to the field mass. (Notice
that, during the (psedo--)stationary expansion it is $\Phi \propto
\ln(a)$ and this moderate growth continue also after WDM --and
baryons-- become dominant.) The fact that this occurs slightly after
the time when the dilution of the kinetic energy of WDM particles
lends relevance to their mass, might not be casual.

This class of cosmologies is however peculiar for the very
(pseudo--)stationary expansion process. The cosmic components keep
steadily fixed proportions, as we delve into earlier and earlier eras.
One might even tentatively guess that the observed distribution was
fixed at the end of inflation. If we tentatively argue that the $\Phi$
field we are perceiving as DE coincides with the inflationary field,
we face a number of problems that we plan to discuss elsewhere.

The above issues concern background features. When we come to
fluctuation dynamics, the first point is that the $C_\ell$ CMB spectra
appear barely indistinguishable from $\Lambda$CDM. A priori this is
not obvious, as coupled CDM fluctuations, on the last scattering band
and --even more-- later on, already significantly exceed WDM, and the
low--$\ell$ region, where fluctuations are essentially due to gravity,
could be influenced.

Let us now come to data fit. The point will be discussed here from a
semi--qualitative side, namely to argue whether: (i) there is any
perspective that S.C. cosmologies may ease the problems met by
$\Lambda$CDM models; (ii) the large--$z$ behavior is substantially
affected.

A first difficulty of $\Lambda$CDM concerns the amount of substructure
in Milky Way sized haloes \cite{klypin-moore}. Models involving CDM
overpredict their abundance by approximately one order of magnitude. A
second issue concerns the density profiles of CDM haloes in
simulations, exhibiting a cuspy behavior \cite{moore1994,
  FloresPrimack1994-Diemand2005-Maccio2007-Springel2008}, while the
density profiles inferred from rotation curves suggest a core like
structure \cite{deBlok2001-KuziodeNaray2009-Oh2011}. A third issue
concerns dwarf galaxies in large voids: recent studies
\cite{Tikhonov2009-Zavala2009-PeeblesNusser2010} re-emphasized that
they are overabundant.

We believe that this class of cosmologies might substantially ease the
hardest of these problems, the only one which seemingly found no
reasonable solution yet: the question of density profiles.  

As a matter of fact, it is known that replacing CDM with a ``warmer''
DM component, as a thermal relic of particles whose mass is $\sim
2$--3 keV, yields predictions better than $\Lambda$CDM.
There is a number of ``thermal'' candidates for such WDM; among them,
a sterile neutrino and a gravitino
\cite{AbazajianKoushiappas2006-Boyarsky2009a} find a reasonable
motivation in particle theory
\cite{DodelsonWidrow1994-Buchmueller2007-TakayamaYamaguchi2000}. 

The free streaming of such particles, in $\Lambda$WDM cosmologies
causes a strong suppression of the power spectrum on galactic and
sub-galactic scales
\cite{Bond1980-BonomettoValdarnini1984-DodelsonWidrow1994-HoganDalcanton2000-ZentnerBullock2003-Viel2005-Abazajian2006,BV},
as we also saw in Figures \ref{TF20} and \ref{TF30}
(green--yellow--black dotted curves). As a consequence, N--body
simulations of these models show a shortage of galactic satellites,
partially easing the observed lack of substructure in the Milky
Way. However, the very Figures \ref{TF20} and \ref{TF30} --as well as,
more clearly, Figure \ref{D2}-- show a spectral gap (up to $\sim
1~$order of magnitude) followed by a power recovery at greater $k$'s,
in the models discussed here. It is then unclear whether and how such
conclusions can be extrapolated to this case.

As far as halo profiles are concerned, they are expected to be similar
to CDM haloes in the outer regions, but flattening towards a constant
value in the inner regions; this was predicted in
\cite{Villaescusa-NavarroDalal2011} and found in simulations
\cite{Colin2008-Maccio2012}. However, the core size found is 30--50
pc, while the observed cores in dwarf galaxies are around the 1000 pc
scale \cite{WalkerPenarrubia2011-JardelGebhard2012}. A dwarf galaxy
core in this scale range would be produced by higher velocity
particles, as those belonging to a thermal distribution if their mass
is $\sim 0.1$--0.4 keV. $\Lambda$WDM cosmologies whose warm component
is made of particles with such a mass, however, yield a greater
streaming length, exceeding the size of fluctuations able to generate
these very dwarf galaxies, in the first place
\cite{MaccioFontanot2010}.

In view of these difficulties, the idea that WDM is accompanied by a
smaller amount of CDM has already been put forward
\cite{brookbaldi,Maccio':2012uh,blennow}. The WDM particle velocities
could then be greater, while a low--mass population is however
produced by the re--infall of later derelativizing WDM particles in
persisting CDM potential holes. This suggestion was put forward quite
indipendently of any particle or cosmic model, although assuming {\it
  ad hoc} a twofold dark matter component does not certainly ease
coincidence problems.

It seems clear that S.C. cosmologies have no apparent difficulty to
explain the observed cores, therefore. On the contrary, to obtain a
fluctuation suppression on the scale of galactic subhalos one needs a
suitable tuning, which might even be insufficient.

It must be however clear that a greater $\Delta^2(k)$ yields a
larger amount of galactic objects on the mass scale
\begin{equation}
M = {4\pi \over 3} \left( 2\pi \over k \right)^3 \rho_{0c}
\Omega_m = 2.88 \times 10^{14} \left(h\, {\rm Mpc}^{-1}
\over k \right)^3\Omega_m\, M_\odot h^{-1}
\end{equation}
(here $\Omega_m = \Omega_b + \Omega_w + \Omega_c$) only in the mass
variance $\sigma^2(M) < 1~.$ For greater $k$'s, on the contrary, it
simply means an earlier formation of the related galactic
systems. Accordingly, the prediction of S.C. cosmologies amounts to
stating an early formation of small mass objects; this prediction
could extend to the Milky Way satellite scale ($k=100\, h$Mpc$^{-1}$
corresponds to a mass $\simeq 7 \times 10^7 M_\odot h^{-1}$), however
keeping their total number at a level similar (or slightly inferior)
to $\Lambda$CDM predictions. A tentative explanation of their
observational scarsity could then be related to a longer lifetime, in
respect to $\Lambda$CDM predictions. In a sense, the satellites we
observe should then be the latest to form, while the older ones have
become dark.

The shift on the formation time is greater at lower mass scales and
cosmic reionization, whose main factor are small galaxies (see, e.g.,
\cite{1209.1387} and references therein), could have occurred a little
earlier. As a matter of fact, no spectral burst on the $10^6$--$10^8
M_\odot h^{-1}$ mass scale is needed to meet observations, but an
increased spectral amplitude on such mass scales does not harm current
expectations.

Another point to be suitably deepened is the formation of early black
holes. This question was recently discussed in \cite{lamastra}, aiming
to exclude DE state equations unable to produce enough of them.  These
cosmologies surely meet current lower limits, but the whole scenario
of early system formation could suffer significant modifications.

A final point concerns the m.s.~velocity of today's (almost) non
relativistic WDM component, reading
\begin{equation}
v^2 \sim T_{o,w}/m_w~.
\end{equation}
Here, $T_{o,w} = T_{o} S$ is the present WDM temperature parameter,
$S$ yielding its ratio with the CMB temperature. For instance, for the
case in Figure \ref{D2}, yielding $S \sim 0.15$, it is $v \sim 0.3
\times 10^{-3}c~,$ a small but non negligible value.

\section{Conclusions}
In this work we aim to show that cosmologies allowing for a strong
energy flow from CDM to DE may be quite promising. DE is treated as a
scalar field $\Phi$, and the energy flow is fixed by a coupling
constant $\beta \gg \sqrt{3}/2$~.  This kind of coupled--DE theories
is not new in cosmology, but such large couplings were excluded, up to
now, as coupled CDM was supposed to be the only DM component; if so,
it must be $\beta <\sim 0.15$.

If we lift the limit on the $\beta$ coupling, however, we find that a
dual component, made of coupled non--relativistic particles and the
scalar field $\Phi$, falls onto an attractor solution, being in
equilibrium with radiative components in the radiative era. More in
detail: the early density parameters converge onto fixed values
\begin{equation}
\label{c2c}
\Omega_c = 1/(2\beta^2) ~~~{\rm and} ~~~ 
\Omega_d = 1/(4\beta^2),
\end{equation} 
for CDM and the field, respectively; such density parameters keep
constant, as both dual components dilute $\propto a^{-4}$ as the
Universe expands. This solution being an attractor, if we set initial
conditions violating eq.~(\ref{c2c}), the densities of CDM and DE
fastly mutate and the condition (\ref{c2c}) is restaured. This
attractor exists only for $\beta > \sqrt{3}/2$~.

If we suppose that this were the actual cosmic component in the early
Universe, an exit from the steady expansion can be caused either by
another uncoupled CDM component, or by the derelativization of a WDM
component. Diluting then $\propto a^{-3}$, the density of such further
component eventually overcomes the densities of the field and CDM.
When this happens, however, also the ratio between these latter
densities and radiation starts to increase, although more slowly.

In any coupled--DE theory, the DE state parameter $w(a) \equiv +1$, in
the early Universe (there are some exceptions, when a self interaction
potential $V(\Phi)$ yields forces so intense to make the DE--CDM
coupling almost negligible). In order that the $\Phi$ field assumes DE
features, $w(a)$ must eventually turn from +1 to $\sim -1$ about a
suitable redshift $z_d$.  {\sl Such $z_d$ can then be easily tuned to
  allow all cosmic components to reach their observational ratios.}
The $w(a)$ transition is expected to occur because of the progressive
dilution of the field kinetic energy, while the field has a steady
increase $\propto \ln(a)$. All that is indipendent from any specific
assumption on the form of the self--interaction potential $V(\Phi)$.

In this paper we considered both the option that uncoupled DM is cold
or warm. The latter option may however lead to a rather attractive
picture, in which scalar field, coupled CDM, and WDM have close steady
density parameters in the early Universe. The main radiative
components, made of photons and neutrinos, would then be significantly
denser just because heated up by the progressive decays of other
particles belonging to the primeval thermal soup, while WDM, coupled
CDM and $\Phi$ decoupled quite early and in the same epoch.

The main topic of this paper, however, is the analysis of density
fluctuation evolution in such cosmologies. This required first an
analysis of initial conditions. The successive evolution was then
treated both with an heuristic 11--eqs. program, able to outline the
essential physical features, and by suitably modifying the public
CMBFAST program.

We find that fluctuation spectra and their evolution strictly resemble
$\Lambda$CDM so that, in first approximation, {\sl the present large
  scale picture is quite similar to $\Lambda$CDM, although microscopic
  quanta have a mass scale $\sim \, $0.1--0.4~keV.} Possible
discrepancies therefore emerge just below the Milky Way scale,
allowing for a natural explanation of flat halo profiles in dwarf
galaxies and, possibly, for the observed shortage of Milky Way
substructure.

Further significant differences from the $\Lambda$CDM scenario are
however expected during the epoch of reionization, when first
structures form.  In particular, the first stars and the cosmic
reionization are expected to occur earlier; also the primeval
back--hole formation is expected to be more effective.

Let us finally remind that the CMB angular spectra, in S.C.
cosmologies, strictly resemble $\Lambda$CDM for the same parameter
choice.

\vskip .4truecm

\noindent
{\bf Acknowledgments.} Stefano Borgani and Matteo Viel are gratefully
thanked for discussions. One of us (S.A.B.) acknowledes the support of
CIFS.

\vskip 2.truecm

\end{document}